\documentclass[12pt,preprint]{aastex}
\begin{document}

\title{Eclipses During the 2010 Eruption \\ of the Recurrent Nova U Scorpii}
\author{Bradley E. Schaefer, Ashley Pagnotta\affil{Department of Physics and Astronomy, Louisiana State University, Baton Rouge, LA 70803}}
\author{Aaron P. LaCluyze, Daniel E. Reichart, Kevin M. Ivarsen, Joshua B. Haislip, Melissa C. Nysewander, Justin P. Moore\affil{Department of Physics and Astronomy, University of North Carolina at Chapel Hill, Chapel Hill, NC}}
\author{Arto Oksanen\affil{Caisey Harlingten Observatory, Caracoles 166, San Pedro de Atacama, Chile}}
\author{Hannah L. Worters, Ramotholo R. Sefako\affil{South African Astronomical Observatory, PO Box 9, Observatory 7935, Cape Town, South Africa}}
\author{Jaco Mentz\affil{Unit for Space Physics, North-West University, Private Bag X6001, Potchefstroom 2520, South Africa}}
\author{Shawn Dvorak, Tomas Gomez, Barbara G. Harris, Arne A. Henden, Thiam Guan Tan, Matthew Templeton\affil{American Association of Variable Star Observers, 49 Bay State Road, Cambridge MA 02138}}
\author{W. H. Allen\affil{Center for Backyard Astrophysics, Vintage Lane Observatory, RD 3, Blenheim, New Zealand}}
\author{Berto Monard, Robert D. Rea, George Roberts, William Stein\affil{Center for Backyard Astrophysics}}
\author{Hiroyuki Maehara\affil{Kwasan Observatory, Kyoto University, Kyoto Japan}}
\author{Thomas Richards\affil{CBA, 8 Diosma Road, Eltham, Victoria, 3095, Australia}}
\author{Chris Stockdale\affil{AAVSO, 8 Matta Drive, Churchill, Victoria, 3842, Australia}}
\author{Tom Krajci\affil{AAVSO \& CBA, P.O. Box 1351, Cloudcroft, New Mexico}}
\author{George Sjoberg\affil{AAVSO, P.O. Box 2825, 39 Soule Avenue, Duxbury MA, 02331-2825}}
\author{Jennie McCormick\affil{Center for Backyard Astrophysics, Farm Cove Observatory, 2/24 Rapallo Place, Farm Cove, Pakuranga, Auckland, New Zealand}}
\author{Mikhail Revnivtsev, Sergei Molkov\affil{Space Research Institute, Moscow, Profsoyuznaya 84/32, Russia}}
\author{Valery Suleimanov\affil{Kazan State University, Astronomy Department, Kremlyovskaya 18, 420008          
Kazan, Russia, and Institute for Astronomy and Astrophysics, Kepler Center for Astro and Particle Physics, Eberhard Karls Universitaet, Sand 1, 72076 Tuebingen, Germany}}
\author{Matthew J. Darnley, Michael F. Bode\affil{Astrophysics Research Institute, Liverpool John Moores University, Birkenhead, CH41 1LD, UK}}
\author{Gerald Handler\affil{Institute of Astronomy, University of Vienna, T\"urkenschanzstr. 17, 1180 Vienna, Austria; and Copernicus Astronomical Center, Bartycka 18, Pl 00-716         
Warsaw, Poland}}
\author{Sebastien Lepine, Michael M. Shara\affil{Department of Astrophysics, Division of Physical Sciences, American Museum of Natural History, Central Park West at 79th Street, New York, NY 10024, USA}}

\begin{abstract}

The eruption of the recurrent nova U Scorpii on 28 January 2010 is now the all-time best observed nova event.  We report 36,776 magnitudes throughout its 67 day eruption, for an average of one measure every 2.6 minutes.  This unique and unprecedented coverage is the first time that a nova has any substantial amount of fast photometry.  With this, two new phenomena have been discovered: the fast flares in the early light curve seen from days 9-15 (which have no proposed explanation) and the optical dips seen out of eclipse from days 41-61 (likely caused by raised rims of the accretion disk occulting the bright inner regions of the disk as seen over specific orbital phases).  The expanding shell and wind cleared enough from days 12-15 so that the inner binary system became visible, resulting in the sudden onset of eclipses and the turn-on of the supersoft X-ray source.  On day 15, a strong asymmetry in the out-of-eclipse light points to the existence of the accretion stream.  The normal optical flickering restarts on day 24.5.  For days 15-26, eclipse mapping shows that the optical source is spherically symmetric with a radius of 4.1 R$_{\odot}$.  For days 26-41, the optical light is coming from a rim-bright disk of radius 3.4 R$_{\odot}$.  For days 41-67, the optical source is a center-bright disk of radius 2.2 R$_{\odot}$.  Throughout the eruption, the colors remain essentially constant.  We present 12 eclipse times during eruption plus five just after the eruption.

\end{abstract}

\section{Introduction}

U Scorpii is a recurrent nova with ten known eruptions: 1863, 1906, 1917, 1936, 1945, 1969, 1979, 1987, 1999 (Schaefer 2010), and just recently in 2010.  The latest eruption was predicted (Schaefer 2005) and the discovery was made independently by B. G. Harris and S. Dvorak (Schaefer et al. 2010a, 2010b, Simonsen \& MacRobert 2010).  The pre-eruption magnitude was V=18.0, the  eruption started on JD 2455224.32$\pm$0.12, and came to a peak at around V=7.5 mag on JD 2455224.69$\pm$0.07 (Schaefer 2010; Schaefer et al. 2010b).  The initial decline of U Sco is the fastest of all known novae, reaching V=14.0 just 12 days after the peak, then leveling off to a plateau lasting for 20 days, followed by another sharp drop (Schaefer et al. 2010d), then another plateau from days 41-54 after peak (Pagnotta et al. 2010), and finally fading back to the quiescent level and behavior 67 days after the eruption began.

U Sco is a total eclipsing system with a period near 1.23 days (Schaefer 1990; 2010; Schaefer \& Ringwald 1995).  The total eclipses appear as a flat interval in the V-band light curve lasting 0.0253$\pm$0.0025 in phase, during which the white dwarf and most of the quiescent accretion disk must be covered by the G-type sub-giant companion star.  U Sco is unique as a total eclipsing system where we know the eclipse ephemeris in advance and we know when the eruption will happen.  These circumstances are very important, as this allows the {\it pre}-eruption orbital period to be measured to high accuracy, so that subsequent measure of the post-eruption orbital period will give the orbital period change caused by the ejected material and an accurate measure of the mass lost by the binary, which is key for the question of whether recurrent novae are progenitors of Type Ia supernovae.  These circumstances are also critical for the measure of the distance to U Sco, where the observed brightness and temperature of the companion star alone during totality allow for a reliable blackbody distance (12$\pm$2 kpc; Schaefer 2010).  Given that most other distance measures to novae are only accurate to roughly a factor of two (Downes \& Duerbeck 2000), this makes the U Sco distance one of the most accurately known for all novae.  The eclipsing nature of U Sco also allows for the unique ability to get eclipse mapping of the brightness distribution of the supersoft X-ray source (SSS) around the white dwarf.  With our advance notice of the eruption, our group was able to organize intensive observing campaigns in the optical (with results being reported in this paper) as well as in the X-ray and ultraviolet (Schaefer et al. 2010d; Osborne et al. 2010).

Eclipses {\it during} eruptions had already been reported for the 1999 eruption (Thoroughgood et al. 2001; Matsumoto et al. 2003).  Our early report of eclipses starting in the first plateau phase appear in Schaefer et al. (2010c).  This paper will report on the eruption eclipses as based on our 36,776 magnitude measures during the 2010 eruption.

\section{Observations}

With our advance notice for the eruption of U Sco, we had in place an organization allowing us to take very detailed observations throughout the entire eruption.  In particular, we have daily and hourly photometry in the X-ray, the ultraviolet, the optical, the near-, and the middle-infrared, plus we have daily and hourly spectroscopy in the X-ray, ultraviolet, optical, and near-infrared.  We have intensive pre-eruption monitoring from 2000 until the hour before the start of the eruption (Schaefer et al. 2010b).  Our unprecedented large number of observations, broad wavelength coverage, and the range of data types make this 2010 eruption of U Sco the all-time best observed nova event.

In this paper, we will only report on our optical photometry taken throughout the eruption.  Largely, our observations were from programs prepared long before the eruption, including networks of observers associated with the American Association of Variable Star Observers (AAVSO) and the Center for Backyard Astrophysics (CBA).  We made no attempt to get any telescope time on large telescopes as they would be essentially useless, because they would saturate the bright stars and the amount of available time would be negligible.  Instead, we obtained many long sets of fast time series photometry with telescopes of apertures from 0.2 to 2.4 meter, with typical cadence ranging between 1 to 4 minutes.  The smaller telescopes were of use only during the fast fall and first plateau phases, while our larger telescopes worked well throughout.  By having many telescopes distributed widely in longitude, we covered a large fraction of each U Sco orbit throughout the entire eruption, despite U Sco being visible for only around one hour from each site at the beginning of the eruption.  (By the end of the eruption, as U Sco moved farther away from the Sun in the morning sky, we could get 6 hour runs from any of our southern sites.)  In the end, we obtained 36,776 magnitudes, with an average of 2.6 minutes between observations throughout the entire 67 days of the eruption.

A full list of observers, sites, telescopes, and filters is given in Table 1.  The observers Pagnotta and Handler both made observations in many filters (BVRIJHK and UBVRI plus Stromgren y respectively), with a full analysis of the light curve shape, colors, and spectral energy distribution (cf. Pagnotta et al. 2010) being reserved for a separate paper.  For the photometry in this paper, we are concentrating on the essentially V-band magnitudes, the UBVRI fast photometry and the eclipse timing, and this is where we have the 36,776 measures.  All of these magnitudes were taken with CCDs.  Roughly three-quarters were made with a Johnson V-band filter or with CCDs running with no filter.  In all cases, these magnitudes were derived using standard aperture photometry on fully processed images as differential photometry with respect to calibrated comparison stars nearby on the same image.  The comparison stars are all well calibrated in all bands through either the sequences published in Henden \& Honeycutt (1997; also available from the AAVSO) or Schaefer (2010).  

This same calibration was applied to the unfiltered images, with the resulting magnitude for U Sco being close to the V-band magnitude system, but with some systematic offset that varied from observer to observer.  U Sco does not change its colors greatly throughout the eruption (Pagnotta et al. 2010), so this systematic offset should be nearly constant for a given observer.  Indeed, we find that we can reconcile every observer to the Johnson V-magnitude system by taking a constant offset.  This offset is determined for each observer by direct comparison with simultaneous observations from a fully calibrated observer.  The uncertainty in these measured offsets is roughly $\pm$0.04 mag.  The offsets for all observers are presented in the last column of Table 1.

We have used the PROMPT 0.41-m telescopes on Cerro Tololo in Chile to obtain quasi-simultaneous UBVRI time series from the start of the plateau until after the end of the eruption.  This allows us to see the specific color variations throughout the decline, the eclipses, and the later dips.  (All other observers only took colors once or twice a night, and it is difficult to reconstruct the fast color changes from such data.  Only the PROMPT data has both colors and fast time series.)  One PROMPT telescope regularly cycled between B, V, R, then I images (all 40 second exposures), while another PROMPT telescope simultaneously took long series of U images (all 80 second exposures).  These were calibrated by differential photometry with respect to nearby stars of known magnitude.  The average U-V and B-V colors are somewhat more red than measured with other detectors, so we think that there must be significant uncorrected color terms associated with having a nonstandard effective bandpass.  Fortunately, such color corrections will be essentially a constant, so this means that the shape of the color curves will be correct.  In all, we have 11,543 PROMPT magnitudes in UBVRI.

We have also made a series of fast photometry after the end of the eruption for the purpose of timing the eclipses.  These series were made with the CTIO 0.9-m, the MDM 2.4-m, and the San Pedro de Atacama 0.5-m telescopes.  Five eclipses were observed with acceptable timing.  The phases of minimum show a substantial and surprising scatter, so the accuracy of the time stamps for the individual images, the time corrections (including the exposure and heliocentric corrections), and best fit times were independently checked in multiple ways by multiple people.  In all, these telescope systems and the full analysis procedure is identical for these post-eruption times as during the eruption and pre-eruption, so we have full confidence in the reported times.

After the end of the first plateau, the nova brightness became faint enough that some of our telescopes had observing cadences (typically one-minute integrations) that produced high statistical noise.  We need to have photometric accuracy better than the size of the significant variations ($\sim 0.1$ mag), and indeed most of our data has statistical error bars of $<0.03$ mag.  The solution for the post-plateau time series with too optimistic cadences is to bin the data together.  This binning is performed as a weighted average.  The result is a substantial reduction in the scatter of the light curve, at a cost of some time resolution.  With our better data, we have never seen variations with time scales of faster than ten minutes or so.  (Just like during the entire eruption, U Sco during quiescence has flickering with variability only on time scales of one hour or longer; Schaefer et al. 2010b.)  So our bin sizes are 0.001, 0.002, 0.004, and 0.008 days (1.4, 2.9, 5.8, and 11.5 minutes respectively).  With this binning, our 36,776 magnitudes are consolidated into 16,995 magnitudes.

The statistical uncertainty for individual observations is often calculated by the photometry software.  In cases where this is not reported, we adopt a typical value of 0.01 mag.  The uncertainties for the binned magnitudes come from the usual propagation of errors for a weighted average.  The uncertainty for almost all the points is between 0.01 and 0.04 mag, with typical values from 0.01 to 0.02 mag.  In the figures in this paper, the error bars are not plotted so as to allow clarity, with the error bars almost always being smaller than the plotted points.

The time associated with each magnitude is halfway between the start and stop time of each image.  These have been converted to Julian Dates, and then the heliocentric corrections have been applied.  For magnitudes binned in time, we take the average time of all input images.

The orbital phase for each observation is taken from our very well determined ephemeris for the eclipse minima during quiescence. This ephemeris for the Heliocentric Julian Date of the middle of the eclipses in quiescence is
\begin{equation}
HJD=2451234.5387(\pm0.0005) + N \times 1.23054695 (\pm0.00000024).
\end{equation}
This is based on 45 eclipse times from 2001 to 2009 (Schaefer 2011), with the uncertainty of the zero phase during the 2010 eruption of less than 2 minutes.  The curvature in the observed O-C curve is consistent with zero and the sudden period change at the time of the eruption is irrelevant for the ephemeris during the eruption, so this ephemeris is applicable during the eruption.  However, the minimum in quiescence is at the phase when the center of light is covered, and the center of light can move between quiescence and the eruption.  Indeed, the eclipse times during the 1999 eruption were significantly offset from the zero phase of the quiescent ephemeris.  With our many eclipse times throughout the 2010 eruption, we will define this phenomenon.

Table 2 lists 7,752 binned magnitudes (not from PROMPT), with the entire table available only in the on-line version.  The columns are (1) the Heliocentric Julian Date of the middle of the exposure or time bin, (2) the orbital phase according to Equation 1, (3) the V-band magnitude with the one-sigma uncertainty, (4) the detrended magnitude as discussed in the next section, and (5) the observer.

Table 3 lists all 8916 binned magnitudes from PROMPT, with the entire table available only in the on-line version.  This listing is strictly by time, with all bands mixed together so that colors can be seen.  The columns are (1) the Heliocentric Julian Date for the middle of the exposure or time bin, (2) the phase of this time from Equation 1, (3) the band, and (4) the observed magnitude and one-sigma error bar.

Table 4 lists 327 post-eruption magnitudes.  The columns are (1) the Heliocentric Julian Date for the middle of the exposure, (2) the orbital phase according to Equation 1, (3) the observing band, (4) the magnitude with its one-sigma error bar, and (5) the observatory.

Figure 1 shows the overall light curve for the eruption.  We see the fast fall (from days 0-12), a transition time interval (days 12-15), the onset of eclipses (around day 15), the first plateau (days 15-32), the fall after the first plateau (days 32-41), the second plateau (days 41-54), the jittery fall after the second plateau (days 54-67), and the end of the eruption with the return to the quiescent level (day 67).  The exact boundaries between these phases are uncertain by 1-2 days.

No other nova eruption has ever had anywhere near as good a light curve.  Indeed, relatively few novae eruptions have even had full coverage from peak to the return to quiescence (Strope et al. 2010).  U Sco not only has {\it complete} coverage, but also we have magnitudes an average of once every 2.6 minutes throughout the eruption.  This is completely unprecedented.  A handful of nova eruptions have had fast photometry in the past, but these have all been just for a few hours each, with such coverage having no real chance to discover any of the various phenomena we observed for U Sco.  Our data set for U Sco is unique and valuable.

\section{Detrended and Phased Light Curves}

To pull out the light curve for the eclipses, we must remove the overall trend visible in Figure 1. To do this, we have established a trend line which essentially runs across the upper envelope of the light curve to avoid the eclipses.  The trendline is a multiply broken line passing through the light curve at phase 0.25, and is presented in Table 5.  The first column gives the Heliocentic Julian Date for each normal point, the second column gives the time since the start of the eruption (i.e., HJD-2455224.32), the third column gives the V-band magnitude at that time, and the fourth column gives a variety of comments for the associated time.  For times between these normal points, the trend line is given by simple linear interpolation.

Magnitudes from this trend line ($V_{trend}$) are then subtracted from the observed magnitudes ($V$) to get the detrended magnitudes ($V-V_{trend}$) which appear in the fourth column of Table 2.  Detrending is important for the timing of eclipses, as an eclipse superposed on a falling light curve will have its time of minimum biased to later time.  Detrending is also important because it allows us to superpose phased light curves from successive orbital periods.

The phased and detrended light curve is constructed by plotting $V-V_{trend}$ versus the orbital phase.  The eruption can be divided up into intervals during which the light curve is largely stationary.  Figures 2-9 present the detrended and phased light curves for each of these intervals.  These figures have identical magnitude scale for easy comparison, and also show the data displayed twice (each point is plotted with its phase as well as with 1.0 plus the phase) to allow the eclipse to be readily visible around phase 1.0 without break.  Figures 2-9 can be used to see the accuracy of the trendline (where the trendline corresponds to $V-V_{trend}=0$ horizontal line) as well as to see the variations from this trendline (from eclipses, flares, dips, and various asymmetries).  The typical variations in these figures (outside of the aperiodic flares and dips) is under 0.1 mag, and this shows that the trendline corrects for the overall decline of the light curve to better than 0.1 mag and that the orbit-to-orbit variations are typically under 0.1 mag.  

For days 0-9 (Figure 2), the steeply falling light curve is flattened out, so we see a nearly constant detrended light  curve.  Some of the scatter could be due to imperfect detrending.  But some of the variations, like the short rise and falls seen around phases 0.02 and 0.33 (duplicated at phases 1.02 and 1.33), are significant and intrinsic to the nova.  The amplitudes are about 0.1 mag with durations of 0.04-0.06 in phase (1.2-2.1 hours), with these two events occuring on days 8-9 (with peaks at HJD 2455232.979 and 2455234.013). 

For days 9-15 (Figure 3), the light curve is in a transition interval, as the fast decline slows to a stop at the start of the first plateau phase.  The light curve displays large amplitude short flares far above the trend line.  Three of these flares have peaks visible (at HJD 2455235.253, 2455236.080, and 2455237.225), all with a peak of 0.5 mag above the trend line, and all with rise or fall times of 0.02-0.04 in phase (0.6-1.2 hours).  The peak times show no correlation with the orbital phase.  The cause of these early flares is currently unknown.  At the time of the flares, the nova shell is optically thick to the central binary system, as shown by the lack of eclipses and supersoft X-ray flux.  On day 10, for an expansion velocity of 5,000 km s$^{-1}$, the shell has a radius of 4 light-hours.  With this, the flares (which must be smaller than, and likely much smaller than, 0.6-1.2 light hours in size) must involve a small fraction of the shell.  So the picture we get is a small region in the shell producing roughly the same luminosity as the rest of the shell, but only for an hour or so.

For days 15-21 (Figure 4), the light curve shows the first part of the plateau phase.  We see a full eclipse plus asymmetric structure outside of the eclipse.  On day 12.0 there is certainly no eclipse, on day 14.5 we have several isolated magnitudes that might be from a low amplitude eclipse, and on day 15.6 there is certainly a well formed eclipse.  So eclipses reappear sometime in the 3 orbit interval from days 12.0-15.6.  The amplitude is 0.6 mag, the total duration is around 0.29 in phase (8.7 hours), and the shape varies somewhat over the interval.  The sudden appearance of eclipses shows us that the nova shell and wind has rapidly become optically thin (or at least translucent) all the way from the inner binary system out to infinity.  The sudden and sharp turn-on of the supersoft X-ray source (SSS) on days 12-14 (Schlegel et al. 2010) also shows that the optical depth to the binary became small at this same time.  This same time is when the early fast decline stops and the plateau phase begins, with the cause likely being that the outer shell continues fading but the revealed inner binary remains roughly constant and provides the light for the plateau.  At phase 0.5 (and phase 1.5 in the figure), we see what looks like a secondary eclipse with an amplitude of 0.20 mag.  This light curve has a striking asymmetry in that the brightness level at  the phase 0.25 elongation is 0.08 mag brighter than at the phase 0.75 elongation.  This requires that some structure in the binary breaks the symmetry of the line between the two stars.  The shell, the two stars, and the wind from the white dwarf will all respect this symmetry from orbit to orbit, so the only apparent way to break the symmetry is with additional material placed to one side of the line connecting the centers of the two stars.  A reasonable source of asymmetric light is the material coming off the companion star as part of the forming accretion stream.  This stream will have relatively little luminosity of its own, but rather will be bright because the hot region around the white dwarf illuminates the inner edge of the accretion stream.  With the accretion stream leading the companion star, its illuminated {\it inner} edge will appear brightest {\it after} the zero phase when the companion eclipses the white dwarf (i.e., around phase 0.25).  The accretion disk and the accretion stream is blown away by the initial eruption (Drake \& Orlando 2010), and the accretion stream will be the first structure to appear.  The stream can be established on a free fall time scale, but will be limited by the nova wind continuing to blow away the falling material until such time as the wind dies enough to allow the stream to penetrate to the region near the white dwarf.  The outer edge of the forming accretion disk can only be established after the stream penetrates to the circularization radius near the white dwarf.  This outer edge will form on the time scale of the orbital period, but will also be limited by the time when the nova wind has declined enough so that the material will not be blown away.  In principle, the outer edge of the accretion disk will be symmetric and so cannot account for the brightening at phase 0.25, however the `hot spot' (where the accretion stream hits the outer disk creating a large structure that would be illuminated from the inner hot region) could provide an asymmetrically placed structure.    We know of no theoretical model of the resumption of the accretion in the face of a nova wind.  We suggest that the observed asymmetry in the light curve is caused by the illumination of the accretion stream (as opposed to the illumination of the hot spot) simply because the stream will form before the hot spot and this corresponds to the earliest time during the eruption.

For days 21-26 (Figure 5), the light curve shows the central time interval of the first plateau phase, with the SSS shining brightly.  The eclipse deepens to 0.80 mag, and the duration might be somewhat shorter (0.25 in phase).  The light curve shows apparent variations in shape.

For days 26-32 (Figure 6), the light curve covers the last part of the first plateau, during which time the SSS is peaking in luminosity.  The eclipse deepens to 1.1 mag, while the total duration increases to ~0.37 in phase.  The secondary eclipse remains prominent.  To have a secondary eclipse while the nova is bright, we must have the companion star greatly brighter than normal, and this can only be due to the illumination of the inner edge of the star by the luminosity from near the white dwarf.  A secondary eclipse also implies something near the white dwarf doing the eclipsing (with the white dwarf being too small to make any significant eclipse), so the occulter must be a just-forming accretion disk, or the optically thick inner parts of the wind being driven off the white dwarf by the SSS.  The brightness at phase 0.75 apparently varies, but at least on occasion is nearly equal to that at phase 0.25.

For days 32-41 (Figure 7), the light curve is in a fast decline from the first plateau, covering a time when the SSS is rapidly turning off.  The eclipse deepens to 1.4 mag, while the total duration remains nearly the same at 0.36 phase.  The eclipse is definitely asymmetric, as the ingress crosses 0.4 mag at phase -0.15 while the egress crosses 0.4 mag at phase +0.08.  This asymmetry could be caused by material in the accretion stream which is in front of the companion star.  This same material would also be illuminated on its inner side and provide extra light around phase 0.25, causing the obvious asymmetry outside of eclipse.  We see that the secondary eclipse (so obvious in previous days) has now vanished.

For days 41-54 (Figure 8), the light curve covers the second plateau.  We see deep and broad dips scattered apparently randomly from phase 0.25 to 0.85.  These dips are a completely new phenomenon for novae.  The bulk of the light is coming from near the white dwarf (as demonstrated by the deep primary eclipse), so the dips can only be eclipses of this source.  The variability in time and phase demonstrates that the eclipses are not associated with the secondary star.  Eclipses that occur at phases of 0.25, 0.55, 0.65, 0.80, and 0.85 can only come from an accretion disk.  The disk was certainly blown away by the initial eruption, so the disk is being re-established as the accretion stream orbits the white dwarf colliding with itself.  The inclination of U Sco is $\sim80-84\degr$ (Thoroughgood et al. 2001; Hachisu et al. 2000a; 2000b), so the line of sight to the white dwarf passes just above the disk; any high spot in the edge of the disk will cause an eclipse of the central source.  Billington et al. (1996) presents evidence for precedence of a complex disk rim profile in a cataclysmic variable.  The chaotic disk edge will have fast-changing collision regions at any azimuth, so the high edges of the disk can produce eclipses that appear and disappear at any orbital phase.  We name this new phenomenon `optical dips' with U Sco being an `optical dipper'.  This name is taken from an analogous phenomenon seen in low mass X-ray binaries that have an inclination of $\sim80\degr$ with X-ray dips being seen in these X-ray dipper systems (Walter et al. 1981; 1982; White \& Swank 1982; Frank et al. 1987; Balman 2009).  Our explanation for the optical dips has a good precedent from the X-ray dips, and there really is no other explanation for how eclipses can occur at such a wide range of phases.

For days 54-67 (Figure 9), the light curve covers the decline from the second plateau until the return to quiescence.  We see that the optical dips continue.  The phases of the three optical dips covered are 0.40, 0.50, and 0.75, with depths of 0.4 mag.  The primary eclipse is deep with amplitude 1.1 mag, fairly symmetric in shape, and has a duration of 0.25 in phase.

For each of these time intervals, we have constructed an average template for $V-V_{trend}$ as a function of phase.  The light curve varies from orbit to orbit, so all we can do is follow along some average or median light curve.  For the time of the optical dips, we have merely indicated the upper envelope of the superposed light curves.  These templates are tabulated in Table 6.  We hope that some future program will use these templates for detailed and definitive eclipse mapping of the optical light.

\section{Color Curves}

The PROMPT data provides a unique opportunity to watch the UBVRI colors change across the decline, the eclipses, and the late dips.  This large amount of UBVRI simultaneous fast photometry is unique amongst all novae events.

The UBVRI magnitudes reported in Table 3 were collected together to form many individual U-V, B-V, V-R, and V-I colors.  Each U, B, R, and I binned magnitude was differenced from the V-band magnitudes averaged over a 0.005 day interval to calculate the colors.  Some of the magnitudes did not have any nearly simultaneous V-band measures, so these resulted in no color value.  In all, we have 978 U-V, 1821 B-V, 1997 V-R, and 1818 V-I colors.

Phased plots of the color curves are presented in Figures 10-12.  Figure 10 shows the U-V and B-V colors for days 15-32 (over the entire first plateau) during which the primary and secondary eclipses are visible.  Figure 11 plots the V-I color as a function of the orbital phase for days 32-41, with fast decline and deep eclipses over this time interval.  Figure 12 shows the V-R color for days 41-54, over which the light curve has deep eclipses and deep dips.

The striking result from the plots and tables is that the color of U Sco largely never changes.  Through plateaus, fast declines, deep eclipses, secondary eclipses, and late dips, the colors remain essentially constant.   We see no secular changes, no eclipses (primary or secondary), and no dips.  The observed scatter is consistent with measurement errors.  The median colors are -0.35 for U-V, +0.38 for B-V, +0.17 for V-R, and +0.35 for V-I.

There is no precedent for this data.  Observationally, we would expect some color changes throughout all this, for example because eclipsing binaries always are hiding different temperature bodies.  No prior theoretical speculations have been made as to the expected color changes.

The fact that the colors are not changing with orbital phase tells us that all of the light (or at least that which dominates) is the same color.  One obvious way for this to happen is to have most of the light coming from scattered light (in the nova wind) from the very hot central source.  With Thompson scattering dominating in the ionized wind, the scattering will not change the color, so all the light is the same color (i.e., that of the central source).  Thus, we would see the same color whether the companion is covering the central part of the wind (at central eclipse), is covering just an outer part of the bright wind region (during the ingress or egress), or is not covering anything (outside of eclipse).  Similarly, for dips produced by raised rims of the reforming accretion disk, the colors would not change for most of the light all has the same color.  Asymmetries in the brightness level outside eclipse due to light reprocessed off the accretion stream do not translate into color differences because the light scattered off the accretion stream is the same color as light from the central source.   The companion star provides some amount of light, with this component of a different color, but the companion is greatly fainter than the central nova wind region so any color change during secondary eclipse will be small.  It will be difficult to place quantitative limits on the fraction of blackbody light (say, from the companion star) without a detailed model of the expected temperature.

\section{Eclipse Mapping of the Optical Light}

In this section, we will make an analysis of the eclipse templates, considering eclipse depths, shape of the eclipse light curves, and contact times.  This eclipse mapping is not as detailed as could be done, but nevertheless will extract the primary properties of the central optical light source.

We adopt the model results of Hachisu et al. (2000a; 2000b), with the binary separation $a=6.87$ R$_{\odot}$, the radius of the assumed-spherical companion star $R_{comp}=2.66$ R$_{\odot}$, and the orbital inclination of $i=80\degr$.  Alternatively, we can chose the parameters from Thoroughgood et al. (2001), with $a=6.5\pm0.4$ R$_{\odot}$, $R_{comp}=2.1\pm0.2$ R$_{\odot}$, and $i=82.7\pm2.9\degr$, and find that our results do not change substantively.

The orbital position of the companion star is given by the position angle $\Theta$, as measured from the inferior conjunction of the secondary star, so that $\Theta$ equals the orbital phase (running 0-1) multiplied by $360\degr$.  On the plane of the sky, the projected separation of the centers of the white dwarf and the companion star is equal to $a[\sin ^2(\Theta) + \cos ^2 (\Theta) \cos ^2 (i)]^{0.5}$.  The minimum separation (when $\Theta=0$) is 1.2 R$_{\odot}$, which is smaller than $R_{comp}$, with the implication that the inner region around the white dwarf is totally eclipsed.  The separation equals $R_{comp}$ when $\Theta=20.5\degr$ (i.e., phase 0.057) for a phase of totality for the white dwarf lasting 3.4 hours.  The relative sizes and phases of the orbit and eclipse are illustrated in Figure 13.

When we have coverage including the starts of ingresses and the endings of egresses, we see moderately sharp edges to the primary eclipse.  This points to the central optical source being fairly sharp edged (as opposed to a source that slowly fades with radial distance).  This main optical light source is centered on the white dwarf (as deduced from the eclipse times) and provides most of the optical luminosity in the system (as demonstrated by the depth of the eclipses).  This inner source will be some combination of the photosphere closely surrounding the white dwarf plus light scattered and emitted by the wind being driven off the white dwarf by the SSS.  Both sources are spherically symmetric, so we will initially assume that the surface brightness of the light source is only a function of the radial distance from the white dwarf.  (The forming accretion disk will present light only near the orbital plane, and this will be included later in the analysis.)  With this, the first and last contact times for the primary eclipse then give the radius of the optical source.  From Table 6, the ingress and egress contact phases range from 0.11 to 0.20 away from minimum in phase.  This corresponds to $\Theta$ values from $40-72\degr$ and separations of the star centers of $4.5-6.5$ R$_{\odot}$.  The radius of the optical emitting region is then varying from roughly 1.8 to 3.9 R$_{\odot}$.  This is a relatively crude measure because the contact times are difficult to define accurately, yet this calculation does set the scale for the size of the optical emitting region.

The simplest eclipse mapping calculates the shape of the eclipse light curve for a perfectly dark companion star passing in front of a light source that is uniform, circular, and sharp-edged.  The predicted light curve has only one free parameter (the radius of the optical emitting region) and can be compared to the average templates in Table 6.  For many of the epochs, the out-of-eclipse brightness changes substantially with phase, so the ingress light curve is normalized to phase 0.75 while the egress light curve is normalized to phase 0.25.  Perhaps surprisingly, this simplest model reproduces the templates for many of the epochs.  For days 15-21, a radius of 4.1 R$_{\odot}$ matches the template on both ingress and egress with an RMS scatter of 0.02 mag.  This case is illustrated in Figure 13.  The same fit with a slightly smaller radius is also good for days 21-26, although the one well-measured last contact appears to have a flare at phase 0.13 on day 22 which raises the template locally.  For days 41-54, a reasonable fit is made for a radius of 3 R$_{\odot}$, although the ingress suggests a 10\% smaller radius while the egress suggests a 10\% larger radius.  For days 54-67, the ingress is matched for a radius of 3.0 R$_{\odot}$, although the egress is not well fit for any radius.  The good success of this simplest model for days 15-21 and 21-26 is encouraging, it suggests that more complex models are not needed, it points to the emitting region being roughly spherical in shape, and it sets a fairly accurate scale for the characteristic size of the emitted light.

Despite the success of this simple model (a uniform, circular, sharp-edged optical emission region), the reality is certainly more complex.  We have light coming from the inner edge of the disk, from the accretion stream, from the companion star, from the spherical nova wind, and from the small clumpy gas clouds produced as `spray' where the accretion stream hits the disk.  Most of this light will be reprocessed from X-ray and ultraviolet photons originally emitted near the white dwarf.  The dominant scattering mechanism is Thompson scattering off the electrons, where 5\%-10\% of the energy produces optical light.  Suleimanov et al. (2003) shows how multiple scattering off many small clouds in the `spray' result in up to half the high energy irradiation being reprocessed to optical light.  When viewed with the poor resolution of eclipse mapping, these complexities produce light curves consistent with the uniform circular model.

The simplest model does not work for days 26-32 and 32-41.  In these cases, the observed light curves fall substantially below (fainter than) the model for phases away from the eclipse middle.  The sense is that there is much optical light coming from regions away from the center, so that the system is relatively faint around phases 0.2 and 0.8.  That U Sco is faint so near to elongation implies that there are substantial amounts of light at distances of 3-4 R$_{\odot}$ from the white dwarf (i.e., just inside the orbit of the companion star).  As demonstrated below, this light is apparently confined to the orbital plane, so a good idea is that this is associated with the proto-accretion disk.

The next step is to examine more complex models, while still keeping the radial symmetry.  One possibility is to add in `extra' light that is never eclipsed.  We have tried to add in uniform disks on top of the simplest model, with some of these approximating a point source in the middle.  Another possibility is to add in a `corona' around the uniform disk, and we have tried uniform and sharp-edged coronae, linearly declining coronae, and coronae with power law declines.  Also, we have tried two-dimensional Gaussian and exponential sources.  In all cases, these radially-symmetric models completely fail to reproduce the observed templates.  (For days 15-21 and 21-26, where the simplest model fits well, we have also examined these cases with more complex models, yet none of them provide any improvement.)  The cases in which the simplest model does not work are not improved by arbitrary radial distribution of the emitted light.

For the day 26-32 template, at zero phase, the companion star covers 66\% of all the optical light (as the template magnitude is 1.16 mag).  At phase 0.86 on the ingress the companion star covers 29\% of the total light, while at phase 0.14 on the egress the companion star covers 17\%.  The positions of the companion star at phases 0.86, 0.00, and 0.14 are completely non-overlapping (see Figure 13), so the three covered fluxes add up to 112\%.  Similarly, for days 32-41, the three covered fluxes add up to 113\%.  That the total flux adds up to more than 100\% is not worrisome because the exact number depends sensitively on what is taken as the baseline brightness level for the template.  The adopted baseline is for phase 0.25, but the flux at other times is roughly 10\% less, so we can view the excess flux as simply being an artifact because one phase happens to have added light that is not visible at other phases.  The important point is that nearly all of the light is accounted for by those three positions along the projected orbital plane, so there can be little light above or below the plane at comparable distances.  That is, the light being eclipsed at phases more than 0.14 from central eclipse cannot be radially symmetric, but rather must be from some source in the orbital plane.  The obvious source is the forming accretion disk, which Hachisu et al. (2000a) models as having expanded in radius out to $R_{disk}=3.1$ R$_{\odot}$.

For days 26-32 and 41-54, we have performed eclipse mapping with the optical light source being in the shape of a disk in the orbital plane.  In particular, we take the disk to have the same tilt as the orbital plane, to be cylindrically symmetric and thin, and to have a power law radial distribution of surface brightness.  For trying to match the light curve templates, this model has only two free parameters: the radius of the outer edge of the disk and the power law index for the surface brightness distribution.  Within this model, we find good matches to the templates for $R_{disk}\approx3.4$ R$_{\odot}$ and a power law index of +1.  There are small deviations, but these are easily attributable to the usual variations seen from orbit to orbit.  The power law index of +1 means that the surface brightness increases linearly with distance from the white dwarf, which is to say that the U Sco disk is brightest towards its rim.  In normal quiescent accretion disks, the surface brightness increases as the center is approached.  But center-bright disks and uniform disks are certainly greatly different from the templates, as they do not provide enough eclipsed light in the wings of the eclipse.  The explanation for the bright rim of the U Sco disk is likely some combination of light scattering off the high rims (with the rims being high due to the chaotic impacts from the accretion stream) and the fact that the inner regions have not had much matter filling them yet.  The success of this disk model gives reasonable confidence that the proto-disk around the white dwarf is dominating the optical light from days 26-32 and 41-54.  This disk is substantially larger than in quiescence (Hachisu et al. 2000b), and indeed comes close to the inner Lagrangian point (with radius $6.87-2.66=4.21$ R$_{\odot}$).

For days 41-54 and 54-67, the shape of the primary eclipse is similar to that during quiescence.  The quiescent eclipse template is closely matched by a disk of radius 2.2 R$_{\odot}$ that is centrally bright.  A nearly identical shape is seen during the ingresses for these two time intervals, except for the uncharacteristic shoulder early in the day 54-67 ingress, which we attribute to an optical dip on just one ingress.  The egresses for the two intervals are also similar to the quiescent egresses, except that they are a little wider.  This points to some small asymmetry somewhere near the outer edge of the disk, such as a bright accretion stream that would add some extra light (outside the disk) so as to be covered after the primary eclipse.  This similarity points to the structure of the inner optical source for the late tail of the eruption being largely the same as for the pre-outburst system.  The lack of a spherical component in the optical light is indicated by the narrower eclipse light curves, as demonstrated by the much smaller FWHM values in Table 7 for times after day 41.  This is not surprising as the SSS wind has already turned off (from days 32-41) so we are left with only the accretion disk, just like the quiescent system.

However, for days 41-61, the out-of-eclipse behavior is greatly different from that during quiescence, with the eruption showing unique  and deep dips.  These dips are illustrated in Figure 14.  In Section 3, we presented the logic for why these optical dips must be eclipses from the edge of the accretion disk, as well as the strong precedent for optical dippers from the class of X-ray dippers.  The brightest region of the disk will be its inner portions, and these can be completely covered by a fairly localized raised rim.  The rapidly changing structure of the raised rim is likely associated with the chaotic collisions of the accretion stream as the disk re-establishes itself.  The cessation of dips after day 61 indicates that the chaos has damped out and the disk has settled into a stable structure.  In principle, the structure of the raised rim can be determined from the changing dips in the light curve, hopefully matched to a physical model of the orbital path of the accretion stream (c.f., Billington et al. 1996).  The dip amplitude is 0.5-0.7 mag, which implies that the rim covers over a third of the total optical light.  To cover the inner portions of the disk, the raised rim must extend an angle of $\gtrsim$10$\degr$ over the plane of the disk.  With total dip durations of order 0.25 in phase, the raised rim extends roughly 90$\degr$ in azimuth around the white dwarf.

A full eclipse mapping would account for the physics of the emitting region as well as for the structures in the accretion stream and the proto-disk that make for the complex variations (both dips and asymmetries) outside of eclipse.  Nevertheless, even the simple eclipse mapping gives a clear picture.  For days 15-21 and 21-26, the optical light source appears as a fairly uniform disk of radius near 4.1 R$_{\odot}$ centered on the white dwarf.  This source is the wind ejected from the white dwarf by the SSS due to continued nuclear burning.  For days 26-32 and 32-41, the optical light source is concentrated towards the orbital plane and extends out to near the inner Lagrangian point, with an acceptable model being a rim brightened disk of radius $\approx3.4$ R$_{\odot}$.  So the accretion disk has already formed with a high rim and not-yet-filled inner regions.  For days 41-54 and 54-67, the eclipse mapping shows that the disk has largely settled down to its normal configuration (with radius around 2.2 R$_{\odot}$ and bright in the center), except that from days 41-61 the chaotically raised rims eclipse the inner part of the disk.

\section{Eclipse Times}

The eclipse times of U Sco are important because they are the key to measuring an accurate and reliable $M_{ejecta}$, which is central to the question of whether the white dwarf is gaining mass and will become a Type Ia supernova.  The times of minimum light will be when the center-of-light is covered, and during quiescence these eclipse times are stable with respect to the times of the conjunction of the white dwarf and the companion star (with a measured jitter of 3.5 minutes).  During eruption, as considered in this paper, the center-of-light is shifting around, and this largely precludes the possibility of using these times for purposes of measuring changes in the orbital period of U Sco (but see below).

An accurate time for an eclipse requires coverage that includes the minimum plus both sides of the minimum.  We have good coverage for twelve eruption eclipses.  We define the eclipse time to be the time of minimum light, which is when the center of light is blocked out.  With the rapidly varying asymmetries during the ingress and egress, we only pay attention to the time interval near the minimum.  In practice, we measure the time of minimum by fitting a parabola to the magnitudes in a tight time interval.  This fit is done as a chi-square minimization, with the one-sigma uncertainty being determined by the times over which the chi-square is within 1.0 of its minimum value.  Figures 15 and 16 provide two illustrations of the eclipse light curve and our parabola fit.  These figures show that the scatter of the individual magnitudes around the best fit parabola is generally larger than the error bars, with the differences being larger than any real observational uncertainty.  So the variation is apparently intrinsic to U Sco, and can only be due to some combination of variability in the source and ingress and egress of fine structure across the face of the eclipsed region.  These fast and small variations cannot be modeled, so we have treated them simply as an additional systematic error added in quadrature to the magnitude's measurement uncertainty.  This systematic uncertainty is varied until the final best fit has a reduced chi-square of near unity.  With this, the formal uncertainties in the time of the minimum become a realistic measure in the presence of the variations as seen in the light curves.  Table 7 provides a list of all our eclipse times and properties.

Figure 17 plots the $O-C$ values as a function of the time since the start of the eruption.  The O-C curve starts out significantly and substantially negative, which implies that the early eruption eclipses come earlier in time than expected based on the quiescent eclipse times.  This behavior is expected due to the shift of the center of light between eruption and quiescence.  During the eruption while most of the optical light is coming from a region centered on the white dwarf, the eclipse minima will be at times when the companion star is at inferior conjunction.  During quiescence, the center of light will have shifted from the white dwarf towards the hot spot (where the accretion stream hits the outer edge of the disk).  For the extreme case where the hot spot is maximally offset around the leading edge of a disk with radius 1.78 R$_{\odot}$ and an separation of 6.87 R$_{\odot}$ between the stars (Hachisu et al. 2000a), the center of light will shift by 0.040 in phase (0.050 days).  The quiescent system will have its center of light somewhere between the hot inner edge of the accretion disk centered on the white dwarf and the extreme hot spot position, and thus eclipse minima will occur perhaps $\sim 0.02$ days {\it after} the conjunction between the star centers.  When compared to the quiescent ephemeris, the eruption eclipse times (with minima at inferior conjunction of the companion star) will occur early, and the O-C values will be negative.

Figure 17 shows a significant systematic variation of the O-C curve, with the values getting more positive with a roughly linear trend.  This can only be caused by shifts in the center of light.  As the O-C becomes more positive, the center of light is moving away from the white dwarf and towards the position of the hot spot.  This shift requires the formation of some asymmetry, which can only be structure in the accretion disk or the accretion stream.  As such, Figure 17 is good evidence for the establishment of such structure by day 23-30.

Figure 17 shows that the O-C is positive for eclipses late in the tail.  This implies that the center of light has moved to {\it outside} the quiescent position of the center of light.  This is expected due to the larger size of the accretion disk during the eruption, 3.06 R$_{\odot}$ (Hachisu et al. 2000a) versus the size during quiescence, 1.78 R$_{\odot}$ (Hachisu et al. 2000b).  Figure 17 shows scatter that is larger than the uncertainty in the measurement errors.  Again, we attribute this to the structure in the accretion disk (the high edges that form the dips are active around this time).  We have further eclipse times on days 110-200, with $O-C$ varying from 0.0000 to +0.0054 days.  This observed scatter is larger than expected, and we are suspicious that the accretion disk is still unsettled with resulting variations compared to the quiescent ephemeris.

The literature contains a number of wrong claims about eruption eclipses.  Kato (1999) claims to identify an eclipse from short time series in days 5-11 of the 1987 eruption, but we reject the report because the time series looks to be random variations, the `eclipse' is greatly too short, and we now know that eclipses cannot happen at that time because the inner binary is still completely shrouded.  Matsumoto et al. (2003) identify a secondary minimum on day 6 of the 1997 eruption, but we reject this claim because the minimum is really just an insignificant inflection in the light curve and because we know that eclipses cannot be visible at such an early time.  Matsumoto et al. (2003) combined their eruption eclipse times with eclipse times from quiescence (Schaefer \& Ringwald 1995) to deduce an orbital period change, but we now know that the systematic shifts in the center of light dominate over any orbital effects, so the claimed period change is wrong.  Similarly, for the recurrent nova CI Aql, Lederle \& Kimeswinger (2003) made the same mistake in claiming an orbital period change based on mixing eruption and quiescent eclipse times.

Nevertheless, the eruption eclipse times can be used to place constraints on the orbital period of U Sco.  If we take a pair of eclipses from 1999 and 2010, with both having the same time since peak, then the offsets should be nearly identical and so the time difference should be exactly an integer number of orbits.  The one good Matsumoto primary eclipse time can be paired with the Oksanen time for 20.6 days after the peak, to get an average period of 1.2305479$\pm$0.0000031 days over the 3243 cycles between.  The other two 1999 eclipse times can be similarly paired to get periods of 1.2305439$\pm$0.0000031 and 1.2305494$\pm$0.0000018 days.  The weighted average period is 1.2305480$\pm$0.0000014 days.  The true uncertainty will be somewhat larger than quoted because the reproducibility of eclipse times (see Figure 17) is somewhat larger than the formal measurement error.  This is very close to the independent period expressed in Equation 1.  Unfortunately, this result does not account for any steady period change (i.e., a parabolic term in the O-C curve, such as expected from steady conservative mass transfer). However, this result places a very tight joint constraint on the period and the period derivative, and this will be important for the measure of the sudden period change across an eruption.

\section{The Mass of the Shell}

A strong motivation for obtaining the eclipse depths throughout the eruption was to use the data to measure the amount of mass ejected in the nova shell.  Let us consider an idealized case where the ejecta is all lost over a short period of time around the start of the eruption.  The mass of this shell will be
\begin{equation}
M_{shell}=4 \pi \int_0^{\infty} \rho r^2 dr,
\end{equation}
where $\rho$ is the density of the shell as a function of radial distance from the white dwarf.  The density will be $\rho = m_H \mu n_e$, where $m_H$ is the mass of the hydrogen atom, $\mu$ is the number of nucleons per electron (1 for pure hydrogen, $\sim$2 for heavier elements), and $n_e$ is the electron density.  At a late time when the the shell has expanded to a radial size much larger than the shell thickness, the `$r$' value in the integral can be approximated as $\langle V \rangle \Delta T$, where $\langle V \rangle$ is some average effective velocity and $\Delta T$ is the time since the ejection of the shell.  The optical depth through this shell is
\begin{equation}
\tau=\sigma_{Th} \int_0^{\infty} n_e dr,
\end{equation}
where $\sigma_{Th}$ is the Thompson cross section appropriate for the dominance of electron scattering in this highly ionized shell.  Now, we have
\begin{equation}
M_{shell}=4 \pi m_H \mu (\langle V \rangle \Delta T)^2(\tau / \sigma_{Th}).
\end{equation}
If at some time, we can determine the optical depth through the shell, then we can easily calculate a reasonably accurate shell mass.

Before the eruption, we had expected to watch the depth of the eclipses change, deepening as the shell cleared due to its geometrical expansion.  With the eclipse depth during quiescence setting the level for what the underlying system is doing, we would use the eclipse depths to determine the optical depths.  The effects of the added light from the shell would have to be subtracted out, and the changes of the brightness of the system components would have to be modeled.  Indeed, we see the eclipse amplitudes increasing from near zero on day 12, to 0.6 mag on day 15.6, to 0.8 mag around day 24, to 1.1 mag around day 29, to 1.4 mag around day 37.  After the eruption, we realized that another way to get the optical depth to the central binary is to use the supersoft X-ray luminosity, which is roughly constant throughout the stages before the end of the first plateau.  In particular, the sudden turn-on of the SSS flux starting around day 12 is caused by the thinning of the material from around the binary.  By either measure, it looks like $\tau \sim 1$ for $\Delta T \sim 14$ days.  With these numbers, we get the completely unreasonable answer of $M_{shell} \sim 0.01$ M$_{\odot}$.

Unfortunately, we now realize that this idealized situation is not relevant for U Sco.  In particular, the initial shell had already expanded to optical thinness by the time of the start of the plateau, so the increase in the eclipse amplitude and the turn-on of the SSS cannot be related to the expansion of the shell.  This can also be seen by realizing that the turn-on of the SSS is much too fast to have been caused by the geometric expansion of a shell ejected 12 days earlier.  With the outer shell being already optically thin, the visibility of the eclipses and SSS are being controlled by the thickness of the wind driven off the white dwarf.  This SSS wind happens to thin out around days 12-16, but this has nothing to do with the mass in the shell.  In principle, we could apply Equation 4 to the SSS wind, but in practice the changes in optical thickness are dominated by the rate of the turn-off rather than by the geometrical dilution from expansion, so we cannot even get a mass ejection rate for the wind.  In all, we are disappointed, but it appears that this idea to measure $M_{shell}$ cannot work in any case that we recognize.

\section{The Onset of Accretion and the Formation of the Accretion Disk}

The initial eruption sends out a shell which entrains the material in the accretion disk, so the disk is blown away on a fast time scale.  Nevertheless, matter keeps falling off the companion star through the inner Lagrangian point.  Indeed, given the initial heating of the atmosphere of the companion star, the mass overflow rate can only increase soon after the start of the eruption, only to later decline back to normal on some unknown time scale.  At first, the resulting accretion stream will be blown away as its material gets swept out by the expanding shell and by the subsequent wind being driven by the SSS.  As the SSS wind weakens, the accretion stream penetrates deeper down towards the white dwarf.  There will be some time interval during which the accretion stream is present but no disk (or circularization ring) is present because the nova wind blows away the closer material.  At some time, the accretion stream will penetrate far enough that the stream can pass around the white dwarf and return back onto itself, which creates an impact region that is enlarged.  On the orbital time scale, the stream will create other collision regions, spread out in azimuth around the white dwarf.  These collisions will circularize the material into a proto-disk.  On the time scale by which viscosity spreads the disk, the full accretion disk will form, with matter being transferred to the inner edge of the disk, to ultimately accrete onto the white dwarf.

The time for the onset of the various phases of the re-establishment of accretion is a complex question, for which we know of no theoretical study.  From an observational point of view, we only know of one result, that being the startup of accretion after the eruptions of RS Oph.  The RS Oph brightness falls quickly (as appropriate for a recurrent nova) to fainter than normal quiescent level (Schaefer 2010), which can only happen because the accretion disk is not contributing its normal amount of light.  This post-eruption dip lasts for typically half a year before the system returns to its normal quiescent brightness, so this must be the time scale for the re-establishment of the accretion disk.  This long time scale must be comparable to the orbital period (457 days).  As the post-eruption dip is ending, the ordinary flickering restarts (Worters et al. 2007), and this flickering is a product of the disk.  This pattern (post-eruption dip and the onset of flickering) is a clear indication of the reformation of the accretion disk.  No other nova has post-eruption dips, although only RS Oph has a $>1$ year orbital period and a short eruption duration so that any such dip could be detected.  ( T CrB has complications caused by its unique post-eruption brightening.  V745 Sco and V3890 Sgr, the two other long-period RNe, have not been followed long enough after eruption to recognize any post-eruption dip.)  Nevertheless, the {\it lack} of post-eruption dips in all other novae provides proof that the time scale for the reformation of the disk is faster than the timescale for the end of the eruption.

For U Sco, the out-of-eclipse light curve starts to show a distinct asymmetry (with the elongation at phase 0.25 being brighter than the elongation at phase 0.75) starting around day 15.  This requires some light source that is not centered on the white dwarf or the companion star.  The only possibility for this asymmetrically placed light source is from the accretion stream.  The accretion stream will lead the companion star, so the hot, illuminated inner edge will produce extra light at phases just after the primary eclipses.  As such, the obvious asymmetry is good evidence that the accretion stream is already in place on day 15 (when the central binary system becomes visible).

The folded light curves from days throughout the first plateau are strikingly similar to those of the eclipsing supersoft sources CAL87 (Schmidtke et al. 1993) and RX J0019.8+2156 (Will \& Barwig 1996).  These systems are similar to U Sco in eruption as both have companion stars with roughly one-day orbital periods, accretion disks with high accretion rates, and very hot white dwarfs providing large luminosities that irradiate the accretion structures.  Detailed modeling of the optical light curves of these supersoft sources (Suleimanov et al. 2003; Meyer-Hofmeister et al. 1997; Schandl et al. 1997) account for the unusual out-of-eclipse asymmetry as arising from the high rim of the accretion disk being higher around the hot spot (where the accretion stream impacts the outer disk), with this high edge both eclipsing the inner regions at some phases and providing extra bright light from its irradiated inner edge.  Part of these models is a `spray' from the hot spot that appears as optically thick, cold, clumpy gas clouds (embedded in a surrounding corona) that lie outside and often above the disk, with these making for partial or full obscuration of the inner regions.  Suleimanov et al. (2003) demonstrate that the supersoft X-rays from near the white dwarf are scattered off multiple clouds in this spray, with a substantial fraction of the impinging energy being reprocessed to optical light.  These analyses for high-inclination supersoft sources appear to apply directly to U Sco once its accretion disk has restarted.

The usual flickering is a certain indication of an accretion disk, even if it is unclear where the flickering arises within the disk.  The early flares (as reported by Worters et al. 2010 and Munari et al. 2010) are {\it not} flickering because they occur at times when the inner binary system is certainly not visible.  The first real flickering (with multiple connected changes both above and below the trend line) is on day 24.5.  So we know that at least some part of the accretion disk has already formed by day 24.5. This newly formed accretion disk can only be the outer part of the disk (including the hot spot where the accretion stream impacts the outer disk) around the circularization radius, because it takes some time before the inner part of the accretion disk can get material by viscosity from the outer edge.  Day 24.5 is just before our eclipse mapping (see Section 5) shows that the disk has formed with only the outer rim being bright.  The demonstrates that the early flickering in U Sco comes from the outer part of the disk, and not the inner part.  There has long been questions about where flickering arises, and in this case we have a clear answer.

The day 20.6 eclipse time is early by 0.0134 days, with a systematic shift as the eruption progresses.  This shift in the center of light can only be due to some extra source of light near the position of the hot spot, and this can only be due to structure in an accretion disk (like the hot spot).  With shifts being seen by day 23-30, we have good evidence that a disk with structure has formed by this time.

From day 26 to 41, the eclipse light curves show very wide wings.  Eclipse mapping shows that there must be a substantial amount of flux coming from a light source flattened into the orbital plane.  Detailed eclipse mapping shows the light source to be a disk with $R_{disk}\approx3.4$ R$_{\odot}$ that is bright around the rim.  This points to the disk not having enough time yet to fill in the inner regions.

Starting on day 41, eclipse mapping shows that the optical light source is disk shaped and bright in the center.  This demonstrates that the inner region of the disk has been filled.  The time scale from day 26 to 41 indicates the longer of the time scale for viscosity to spread material form the outer to inner parts of the disk and the time scale for the nova wind to turn off and allow the inward traveling material to not be blown away.

The optical dips start on day 41 and are last seen on day 61.  With most of the optical light coming from near the white dwarf (as shown by the deep primary eclipse), optical dips can only be caused by an eclipse of the innermost region.  The only possible position for an eclipsing body for phases 0.25-0.75 is in the accretion disk.  Based on the strong precedent of the X-ray dippers plus the $\sim80-84\degr$ inclination of U Sco, we have the interpretation of the optical dips as being from high spots on the edge of the accretion disk, with the time variability caused by the chaotic impact of the accretion stream as the disk forms.  So the optical dips are strong evidence that the accretion disk has formed all around the white dwarf by day 41.

The end of the eruption on day 67 has U Sco returning to the normal quiescent level of V=18.0.  This level has the light dominated by the full accretion disk.  The lack of a post-eruption dip proves that the full accretion disk has already formed by day 67.

For the first time, we can measure the times and details for the various phases of the restart of the accretion.  The accretion stream is already present and prominent when the shell clears on day 15.  The outer part of the accretion disk has formed by around day 25, with a large radius and little matter in the inner disk.  By day 41, the disk has largely relaxed to its quiescent configuration.  The impacts of the accretion stream on the proto-disk results in raised rims that create the optical dips, with these instabilities dying away by day 61. 

\section{A Detailed Picture of the U Sco Eruption}

The 2010 eruption of U Sco is the all-time best observed nova eruption, and we here report on 36,776 magnitudes spread over the 67 days of the eruption at an average rate of one every 2.6 minutes.  This exhaustive coverage is unique, and the first time that a nova has had more than a few hours of fast time series photometry.  This unprecedented coverage has allowed for the discovery of two new phenomena: the early fast flares and the late optical dips.  In addition, our analysis measures the nature and timing of the central optical source and the resumption of the accretion.

For days 0-9, U Sco has a fairly smooth fading light curve.  The rate of fade in U Sco is faster than any other known nova, with $t_2=1.7$ days and $t_3=3.6$ days.  Recurrent novae must have fast decline rates, because they must have white dwarfs fairly near to the Chandrasekhar mass, which necessarily implies a small $t_3$ (Hachisu \& Kato 2010).  U Sco is the fastest of the recurrent novae, which points to its white dwarf likely being very near the Chandrasekhar limit, with the possibility that it will soon (on astronomical time scales) become a Type Ia supernova.

For days 9-15, the fading continues, with the outer regions of the shell having expanded enough so that geometrical dilution allows the inner regions of the shell to be visible.  Deep inside the shell, the white dwarf is still undergoing nuclear burning, which drives off a dense wind.  During this time interval, the fast photometry shows the unprecedented and unpredicted phenomenon of the short duration flares.  These flares have no proposed mechanism, although it is clear that this unknown mechanism cannot involve the inner binary (e.g., the accretion disk) as this is still shrouded by the SSS wind.

Around day 15, U Sco is in a transitional phase as the shell and wind clear enough so that the inner binary becomes visible.  The eclipses suddenly appear as the shroud lifts.  The supersoft X-ray photons suddenly can get out as the wind thins.  The inner optical source produced by the wind becomes visible and holds roughly steady (because the nuclear burning is roughly steady), and this steady light source provides a nearly constant flux for the system (despite the continuing fading of the nova shell), which appears in the light curve as the first plateau.

For days 15-21, the light curve shows shallow eclipses and phase 0.25-0.75 asymmetries.  The eclipse mapping shows that the central optical source appears roughly as a uniform disk with a radius of 4.1 R$_{\odot}$, with no extra light.  The shallowness of the eclipse is not caused by the binary still being embedded inside a translucent wind.  Rather, the depth is simply the fraction of the central source covered by the companion star.  The out-of-eclipse asymmetry is visible as soon as the wind clears, and can only be caused by material off the axis connecting the two star centers.  The obvious explanation is that the accretion stream ahead of the companion star will be illuminated by the central source so its inner edge will appear bright to Earthbound observers around phase 0.25.  Thus the accretion stream has established itself by day 15 or earlier.

For days 21-26, the eclipse deepens because of a slight decrease in the size of the optical emitting region (as seen with eclipse mapping).  The asymmetry continues, pointing to the accretion stream which might appear partially embedded inside the translucent luminous wind. The stream interacts with itself, creating a proto-disk, which is the source of the flickering first seen on day 24.5.  This disk will have raised rims, but the optical light comes from such a large source (as eclipse mapping shows) that simple raised rims will occult little light and no optical dips can be observed.  A shallow secondary eclipse is apparent, and this can only be caused by the proto-disk and the inner opaque regions of the wind occulting the bright, irradiated companion star.  By day 26, we see various evidence that the accretion has progressed from a simple accretion stream to a proto-disk.

For days 26-32, the eclipse mapping shows that the central optical light source has switched from a symmetric circle (caused by the wind) to a narrow band along the orbit extending out to near the inner Lagrangian point (caused by the proto-disk).  This proto-disk is rim-brightened, indicative that the inner disk has yet little material.  Again, with little light from the inner regions, the raised rims of the proto-disk cannot create any significant optical dips.

For days 32-41, the optical light source remains in a bright-rim large-radius configuration.  The phase 0.25-0.75 asymmetry still points to off-axis mass such as the illuminated inner edge of the accretion stream.

For days 41-54, we have two developments, both defined by eclipse mapping.  First, the rim-bright proto-disk settles down to the quiescent configuration with a bright center and a smaller disk radius, which indicates that the time scale for material to fill the inner disk is roughly 41 days.  Second, raised rims of the outer disk are seen to eclipse the inner disk, creating optical dips.  These raised rims are caused by the various impact regions of the accretion stream, and were present since around day 21-26, but the optical dips only become apparent around day 41 when the central optical source becomes small.

For days 54-67, the light curve looks like the quiescent light curve, except that optical dips appear before day 61.  The disk has settled down to its normal configuration, and the accretion stream has settled into a stable configuration.  By day 67, the brightness level and phased light curve are back to quiescence.  The complete eruption goes from quiescence to peak in $\sim$6 hours and then back to quiescence in 67 days, making this the shortest known nova eruption.

In this paper, we have provided some detailed analysis that pulls out the primary properties, yet a full physics analysis is still needed.  An example of a good physics analysis for a prior eruption is the analysis of Hachisu et al. (2000a).  But now, with our wonderfully detailed light curves, the analysis can be greatly improved and given fine time resolution.  For example, we can imagine a full hydrodynamical calculation following the accretion stream, its many impacts, and the circularization of material so as to successfully {\it predict} the occurrences and phases of optical dips along with their time evolution.  To this end, we have provided a table of all magnitudes to allow full and independent analyses.  Even though our paper has laid out the detailed configurations of the U Sco eruption in unprecedented detail, we consider this paper a challenge for modelers to step up and provide a full physical model. 

~
~

This work is supported under a grant from the National Science Foundation (AST 0708079).  The Liverpool Telescope is operated on the island of La Palma by Liverpool John Moores University in the Spanish Observatorio del Roque de los Muchachos of the Instituto de Astrofisica de Canarias with financial support from the UK Science and Technology Facilities Council.  MR is thankful for the support of RFBR grant 10-02-00492.  VS is thankful for the support of RFBR grant 09-02-97013-p-povolzhe.  This paper uses observations made at the South African Astronomical Observatory (SAAO).

{}

\begin{deluxetable}{lllllll}
\rotate
\tabletypesize{\scriptsize}
\tablecaption{Photometry Observers and Sites
\label{tbl1}}
\tablewidth{0pt}
\tablehead{
\colhead{Observer}   &
\colhead{Site}   &
\colhead{Telescope}   &
\colhead{Filters}   &
\colhead{Start-Stop (HJD-2455000)}   &
\colhead{$N_{mags}$}  &
\colhead{Offset (mag)}
}
\startdata

Harris	&	New Smyrna Beach, Florida	&	0.4-m Schmidt-Cass.	&	V	&	224.9	-	283.9	&	87	&	0.05	\\
Dvorak	&	Clermont, Florida	&	0.25-m Schmidt-Cass.	&	V	&	224.9	-	259.9	&	424	&	-0.15	\\
Revnivtsev	&	Earth orbit	&	INTEGRAL OMC	&	V	&	225.4	-	234.4	&	133	&	0.00	\\
Handler	&	Sutherland, South Africa	&	0.5-m SAAO	&	V (and UBRIby)	&	225.5	-	249.5	&	55	&	0.00	\\
Worters	&	Sutherland, South Africa	&	1.0-m SAAO	&	V	&	225.6	-	236.6	&	2650	&	0.00	\\
Munari\tablenotemark{a}	&	Verona, Italy	&	0.3-m Schmidt-Cass.	&	V (and BRI)	&	225.6	-	264.6	&	12	&	0.00	\\
Pagnotta	&	Cerro Tololo, Chile	&	SMARTS 1.3-m	&	V (and BRIJHK)	&	225.8	-	295.7	&	59	&	0.00	\\
Stockdale	&	Hazelwood, Australia	&	0.28-m Schmidt-Cass.	&	V	&	226.2	-	268.3	&	2640	&	0.00	\\
Tan	&	Perth, Australia	&	0.24-m Schmidt-Cass.	&	V	&	226.3	-	263.3	&	337	&	-0.07	\\
Stein	&	Las Cruces, New Mexico	&	0.35-m Schmidt-Cass.	&	V	&	226.9	-	261.0	&	1968	&	0.07	\\
Krajci \& Henden	&	Cloudcroft, New Mexico	&	0.28-m Schmidt-Cass.	&	V	&	226.9	-	292.0	&	1284	&	0.00	\\
Monard	&	Pretoria, South Africa	&	0.3-m Schmidt-Cass.	&	Unfiltered	&	228.5	-	228.6	&	279	&	0.00	\\
Sjoberg	&	Mayhill, New Mexico	&	0.36-m Schmidt-Cass.	&	BV	&	229.9	-	262.9	&	665	&	-0.03	\\
Maehara	&	Kyoto, Japan	&	0.25-m Kwasan Obs.	&	Unfiltered	&	231.3	-	275.2	&	1198	&	0.00	\\
Allen	&	Blenheim, New Zealand	&	0.4-m Cassegrain	&	Unfiltered	&	234.1	-	260.2	&	1396	&	-0.45	\\
Richards	&	Melbourne, Australia	&	0.4-m Ritchey-Chretien	&	Unfiltered	&	235.2	-	259.2	&	1086	&	-0.45	\\
Gomez	&	Madrid, Spain	&	0.2-m Newtonian	&	V \& unfiltered	&	237.7	-	247.7	&	65	&	0.00 \& -0.33	\\
LaCluyze \& Reichart	&	Cerro Tololo, Chile	&	Two PROMPT 0.41-m	&	UBVRI	&	240.8	-	292.8	&	11543	&	0.00	\\
Krajci	&	Cloudcroft, New Mexico	&	0.35-m Schmidt-Cass.	&	VR	&	241.9	-	296.0	&	2078	&	0.25 \& -0.50	\\
Oksanen	&	San Pedro de Atacama, Chile	&	0.5-m Cass.	&	Unfiltered	&	243.7	-	424.7	&	7065	&	-0.32	\\
Roberts	&	Harrison, Arkansas	&	0.4-m Schmidt-Cass.	&	V	&	243.8	-	255.0	&	746	&	-0.01	\\
McCormick	&	Aukland, New Zealand	&	0.35-m Schmidt-Cass.	&	Unfiltered	&	248.1	-	249.1	&	300	&	-0.30	\\
Mentz	&	Sutherland, South Africa	&	1.0-m SAAO	&	V	&	248.5	-	248.6	&	190	&	0.00	\\
Darnley	&	La Palma, Canary Islands	&	2.0-m Liverpool Telescope	&	V	&	252.6	-	259.7	&	91	&	0.00	\\
Sefako	&	Sutherland, South Africa	&	1.0-m SAAO	&	V	&	259.5	-	259.6	&	137	&	0.00	\\
Rea	&	Nelson, New Zealand	&	0.3-m Schmidt-Cass.	&	Unfiltered	&	260.1	-	260.2	&	120	&	0.00	\\
Schaefer	&	Cerro Tololo, Chile	&	0.9-m SMARTS	&	I	&	334.6	-	387.7	&	119	&	0.00	\\
Lepine \& Shara	&	Kitt Peak, Arizona	&	2.4-m MDM	&	I	&	382.6	-	382.7	&	49	&	0.00	\\
\enddata
\tablenotetext{a}{Munari et al. 2010}
\end{deluxetable}

\begin{deluxetable}{lllll}
\tabletypesize{\scriptsize}
\tablecaption{Observed V-band Magnitudes for U Sco Eruption
\label{tbl2}}
\tablewidth{0pt}
\tablehead{
\colhead{HJD}   &
\colhead{Phase}   &
\colhead{V (mag)}   &
\colhead{$V-V_{trend}$}   &
\colhead{Observer}
}
\startdata

2455224.9325	&	0.7806	&	7.85	$\pm$	0.10	&	0.09	&	Harris	\\
2455224.9362	&	0.7836	&	8.02	$\pm$	0.10	&	0.25	&	Harris	\\
2455224.9732	&	0.8137	&	7.83	$\pm$	0.01	&	-0.15	&	Dvorak	\\
2455224.9745	&	0.8147	&	7.83	$\pm$	0.01	&	-0.15	&	Dvorak	\\
2455224.9757	&	0.8157	&	7.84	$\pm$	0.01	&	-0.15	&	Dvorak	\\
2455224.9783	&	0.8178	&	7.84	$\pm$	0.01	&	-0.14	&	Dvorak	\\
2455224.9806	&	0.8197	&	7.85	$\pm$	0.01	&	-0.14	&	Dvorak	\\
2455224.9818	&	0.8206	&	7.84	$\pm$	0.01	&	-0.15	&	Dvorak	\\
2455224.9837	&	0.8222	&	7.84	$\pm$	0.01	&	-0.16	&	Dvorak	\\
2455225.5727	&	0.3009	&	8.88	$\pm$	0.02	&	0.07	&	Handler	\\
2455225.5885	&	0.3137	&	8.90	$\pm$	0.02	&	0.07	&	Handler	\\
2455225.6118	&	0.3327	&	8.96	$\pm$	0.02	&	0.09	&	Handler	\\

\enddata
\end{deluxetable}

\begin{deluxetable}{lllll}
\tabletypesize{\scriptsize}
\tablecaption{UBVRI fast photometry from PROMPT
\label{tbl3}}
\tablewidth{0pt}
\tablehead{
\colhead{HJD}   &
\colhead{Phase}   &
\colhead{Band}   &
\colhead{Magnitude}
}
\startdata

2455240.7655	&	0.6473	&	B	&	14.66	$\pm$	0.15	\\
2455240.7661	&	0.6478	&	V	&	14.24	$\pm$	0.08	\\
2455240.7667	&	0.6482	&	R	&	13.90	$\pm$	0.05	\\
2455240.7673	&	0.6487	&	I	&	13.77	$\pm$	0.06	\\
2455240.7679	&	0.6492	&	B	&	14.51	$\pm$	0.21	\\
2455240.7685	&	0.6497	&	V	&	14.24	$\pm$	0.10	\\
2455240.7691	&	0.6502	&	R	&	13.99	$\pm$	0.06	\\
2455240.7696	&	0.6506	&	I	&	13.77	$\pm$	0.03	\\
2455240.7702	&	0.6511	&	B	&	14.44	$\pm$	0.10	\\
2455240.7708	&	0.6515	&	U	&	13.86	$\pm$	0.04	\\
2455240.7708	&	0.6516	&	V	&	14.23	$\pm$	0.03	\\
2455240.7714	&	0.6521	&	R	&	13.97	$\pm$	0.02	\\

\enddata
\end{deluxetable}

\begin{deluxetable}{lllll}
\tabletypesize{\scriptsize}
\tablecaption{Post-eruption eclipse time series
\label{tbl4}}
\tablewidth{0pt}
\tablehead{
\colhead{HJD}   &
\colhead{Phase}   &
\colhead{Band}   &
\colhead{Magnitude}   &
\colhead{Telescope}
}
\startdata

2455334.6462	&	0.9391	&	I	&	17.55	$\pm$	0.02	&	CTIO 0.9-m	\\
2455334.6509	&	0.9429	&	I	&	17.63	$\pm$	0.03	&	CTIO 0.9-m	\\
2455334.6552	&	0.9464	&	I	&	17.66	$\pm$	0.03	&	CTIO 0.9-m	\\
2455334.6595	&	0.9499	&	I	&	17.69	$\pm$	0.03	&	CTIO 0.9-m	\\
2455334.6638	&	0.9534	&	I	&	17.71	$\pm$	0.03	&	CTIO 0.9-m	\\
2455334.6681	&	0.9569	&	I	&	17.81	$\pm$	0.03	&	CTIO 0.9-m	\\
2455334.6724	&	0.9604	&	I	&	17.80	$\pm$	0.03	&	CTIO 0.9-m	\\
2455334.6768	&	0.9639	&	I	&	17.81	$\pm$	0.03	&	CTIO 0.9-m	\\
2455334.6811	&	0.9674	&	I	&	17.87	$\pm$	0.03	&	CTIO 0.9-m	\\
2455334.6854	&	0.9709	&	I	&	17.96	$\pm$	0.04	&	CTIO 0.9-m	\\
2455334.6897	&	0.9744	&	I	&	17.89	$\pm$	0.04	&	CTIO 0.9-m	\\
2455334.6969	&	0.9803	&	I	&	17.98	$\pm$	0.05	&	CTIO 0.9-m	\\
2455334.7012	&	0.9838	&	I	&	18.12	$\pm$	0.06	&	CTIO 0.9-m	\\
2455334.7055	&	0.9873	&	I	&	18.02	$\pm$	0.05	&	CTIO 0.9-m	\\
2455334.7098	&	0.9908	&	I	&	18.11	$\pm$	0.06	&	CTIO 0.9-m	\\
2455334.7141	&	0.9943	&	I	&	18.17	$\pm$	0.06	&	CTIO 0.9-m	\\
2455334.7184	&	0.9978	&	I	&	18.07	$\pm$	0.05	&	CTIO 0.9-m	\\
2455334.7227	&	0.0013	&	I	&	18.16	$\pm$	0.06	&	CTIO 0.9-m	\\
2455334.7271	&	0.0048	&	I	&	18.13	$\pm$	0.06	&	CTIO 0.9-m	\\
2455334.7314	&	0.0083	&	I	&	18.05	$\pm$	0.06	&	CTIO 0.9-m	\\
2455334.7357	&	0.0118	&	I	&	18.15	$\pm$	0.07	&	CTIO 0.9-m	\\
2455334.7402	&	0.0155	&	I	&	18.06	$\pm$	0.07	&	CTIO 0.9-m	\\
2455334.7445	&	0.0190	&	I	&	18.07	$\pm$	0.07	&	CTIO 0.9-m	\\
2455334.7489	&	0.0225	&	I	&	17.98	$\pm$	0.03	&	CTIO 0.9-m	\\
2455334.7532	&	0.0260	&	I	&	17.98	$\pm$	0.03	&	CTIO 0.9-m	\\
2455334.7575	&	0.0295	&	I	&	17.93	$\pm$	0.03	&	CTIO 0.9-m	\\
2455334.7618	&	0.0330	&	I	&	17.82	$\pm$	0.03	&	CTIO 0.9-m	\\
2455334.7661	&	0.0365	&	I	&	17.84	$\pm$	0.03	&	CTIO 0.9-m	\\
2455334.7704	&	0.0400	&	I	&	17.81	$\pm$	0.03	&	CTIO 0.9-m	\\
2455334.7747	&	0.0435	&	I	&	17.79	$\pm$	0.03	&	CTIO 0.9-m	\\
2455334.7790	&	0.0471	&	I	&	17.74	$\pm$	0.02	&	CTIO 0.9-m	\\
2455334.7836	&	0.0508	&	I	&	17.69	$\pm$	0.02	&	CTIO 0.9-m	\\
2455334.7879	&	0.0543	&	I	&	17.64	$\pm$	0.02	&	CTIO 0.9-m	\\
2455334.7923	&	0.0578	&	I	&	17.58	$\pm$	0.03	&	CTIO 0.9-m	\\
2455334.7966	&	0.0613	&	I	&	17.50	$\pm$	0.03	&	CTIO 0.9-m	\\
2455334.8009	&	0.0648	&	I	&	17.50	$\pm$	0.03	&	CTIO 0.9-m	\\
2455334.8052	&	0.0683	&	I	&	17.45	$\pm$	0.02	&	CTIO 0.9-m	\\
2455334.8095	&	0.0718	&	I	&	17.42	$\pm$	0.02	&	CTIO 0.9-m	\\
2455334.8138	&	0.0753	&	I	&	17.35	$\pm$	0.02	&	CTIO 0.9-m	\\
2455376.4857	&	0.9398	&	I	&	17.79	$\pm$	0.03	&	CTIO 0.9-m	\\
2455376.4900	&	0.9433	&	I	&	17.80	$\pm$	0.03	&	CTIO 0.9-m	\\
2455376.4943	&	0.9468	&	I	&	17.81	$\pm$	0.05	&	CTIO 0.9-m	\\
2455376.4986	&	0.9503	&	I	&	17.94	$\pm$	0.05	&	CTIO 0.9-m	\\
2455376.5029	&	0.9538	&	I	&	17.77	$\pm$	0.06	&	CTIO 0.9-m	\\
2455376.5072	&	0.9573	&	I	&	18.05	$\pm$	0.08	&	CTIO 0.9-m	\\
2455376.5115	&	0.9608	&	I	&	17.74	$\pm$	0.08	&	CTIO 0.9-m	\\
2455376.5158	&	0.9643	&	I	&	17.95	$\pm$	0.11	&	CTIO 0.9-m	\\
2455376.5201	&	0.9678	&	I	&	17.93	$\pm$	0.07	&	CTIO 0.9-m	\\
2455376.5244	&	0.9713	&	I	&	18.02	$\pm$	0.07	&	CTIO 0.9-m	\\
2455376.5290	&	0.9750	&	I	&	18.08	$\pm$	0.11	&	CTIO 0.9-m	\\
2455376.5333	&	0.9785	&	I	&	18.03	$\pm$	0.08	&	CTIO 0.9-m	\\
2455376.5376	&	0.9820	&	I	&	17.96	$\pm$	0.09	&	CTIO 0.9-m	\\
2455376.5419	&	0.9855	&	I	&	18.11	$\pm$	0.09	&	CTIO 0.9-m	\\
2455376.5462	&	0.9890	&	I	&	18.35	$\pm$	0.12	&	CTIO 0.9-m	\\
2455376.5505	&	0.9925	&	I	&	18.24	$\pm$	0.08	&	CTIO 0.9-m	\\
2455376.5548	&	0.9960	&	I	&	18.13	$\pm$	0.05	&	CTIO 0.9-m	\\
2455376.5591	&	0.9995	&	I	&	18.24	$\pm$	0.07	&	CTIO 0.9-m	\\
2455376.5634	&	0.0029	&	I	&	18.28	$\pm$	0.08	&	CTIO 0.9-m	\\
2455376.5677	&	0.0064	&	I	&	18.25	$\pm$	0.07	&	CTIO 0.9-m	\\
2455376.5722	&	0.0101	&	I	&	18.29	$\pm$	0.07	&	CTIO 0.9-m	\\
2455376.5765	&	0.0136	&	I	&	18.12	$\pm$	0.06	&	CTIO 0.9-m	\\
2455376.5808	&	0.0171	&	I	&	18.23	$\pm$	0.06	&	CTIO 0.9-m	\\
2455376.5851	&	0.0206	&	I	&	18.16	$\pm$	0.05	&	CTIO 0.9-m	\\
2455376.5894	&	0.0241	&	I	&	18.22	$\pm$	0.05	&	CTIO 0.9-m	\\
2455376.5937	&	0.0276	&	I	&	18.18	$\pm$	0.05	&	CTIO 0.9-m	\\
2455376.5980	&	0.0311	&	I	&	18.09	$\pm$	0.04	&	CTIO 0.9-m	\\
2455376.6023	&	0.0346	&	I	&	18.10	$\pm$	0.04	&	CTIO 0.9-m	\\
2455376.6066	&	0.0381	&	I	&	18.08	$\pm$	0.04	&	CTIO 0.9-m	\\
2455376.6109	&	0.0415	&	I	&	18.01	$\pm$	0.04	&	CTIO 0.9-m	\\
2455376.6157	&	0.0455	&	I	&	17.93	$\pm$	0.04	&	CTIO 0.9-m	\\
2455376.6200	&	0.0490	&	I	&	17.94	$\pm$	0.04	&	CTIO 0.9-m	\\
2455376.6243	&	0.0525	&	I	&	17.80	$\pm$	0.04	&	CTIO 0.9-m	\\
2455376.6286	&	0.0560	&	I	&	17.78	$\pm$	0.04	&	CTIO 0.9-m	\\
2455376.6329	&	0.0595	&	I	&	17.77	$\pm$	0.04	&	CTIO 0.9-m	\\
2455376.6372	&	0.0630	&	I	&	17.63	$\pm$	0.04	&	CTIO 0.9-m	\\
2455376.6415	&	0.0664	&	I	&	17.59	$\pm$	0.03	&	CTIO 0.9-m	\\
2455376.6458	&	0.0699	&	I	&	17.54	$\pm$	0.03	&	CTIO 0.9-m	\\
2455376.6501	&	0.0734	&	I	&	17.54	$\pm$	0.03	&	CTIO 0.9-m	\\
2455376.6544	&	0.0769	&	I	&	17.53	$\pm$	0.03	&	CTIO 0.9-m	\\
2455382.6724	&	0.9674	&	I	&	18.22	$\pm$	0.02	&	MDM 2.4-m	\\
2455382.6755	&	0.9699	&	I	&	18.20	$\pm$	0.03	&	MDM 2.4-m	\\
2455382.6774	&	0.9715	&	I	&	18.24	$\pm$	0.03	&	MDM 2.4-m	\\
2455382.6794	&	0.9732	&	I	&	18.19	$\pm$	0.02	&	MDM 2.4-m	\\
2455382.6814	&	0.9748	&	I	&	18.24	$\pm$	0.02	&	MDM 2.4-m	\\
2455382.6834	&	0.9764	&	I	&	18.26	$\pm$	0.02	&	MDM 2.4-m	\\
2455382.6853	&	0.9780	&	I	&	18.26	$\pm$	0.02	&	MDM 2.4-m	\\
2455382.6873	&	0.9796	&	I	&	18.29	$\pm$	0.03	&	MDM 2.4-m	\\
2455382.6893	&	0.9812	&	I	&	18.27	$\pm$	0.03	&	MDM 2.4-m	\\
2455382.6933	&	0.9845	&	I	&	18.27	$\pm$	0.03	&	MDM 2.4-m	\\
2455382.6953	&	0.9861	&	I	&	18.32	$\pm$	0.02	&	MDM 2.4-m	\\
2455382.6973	&	0.9877	&	I	&	18.33	$\pm$	0.02	&	MDM 2.4-m	\\
2455382.6993	&	0.9893	&	I	&	18.38	$\pm$	0.03	&	MDM 2.4-m	\\
2455382.7012	&	0.9909	&	I	&	18.38	$\pm$	0.03	&	MDM 2.4-m	\\
2455382.7032	&	0.9925	&	I	&	18.36	$\pm$	0.03	&	MDM 2.4-m	\\
2455382.7052	&	0.9941	&	I	&	18.41	$\pm$	0.03	&	MDM 2.4-m	\\
2455382.7072	&	0.9957	&	I	&	18.36	$\pm$	0.02	&	MDM 2.4-m	\\
2455382.7132	&	0.0006	&	I	&	18.41	$\pm$	0.03	&	MDM 2.4-m	\\
2455382.7169	&	0.0036	&	I	&	18.42	$\pm$	0.03	&	MDM 2.4-m	\\
2455382.7189	&	0.0052	&	I	&	18.38	$\pm$	0.02	&	MDM 2.4-m	\\
2455382.7208	&	0.0068	&	I	&	18.36	$\pm$	0.02	&	MDM 2.4-m	\\
2455382.7228	&	0.0084	&	I	&	18.38	$\pm$	0.02	&	MDM 2.4-m	\\
2455382.7248	&	0.0100	&	I	&	18.34	$\pm$	0.02	&	MDM 2.4-m	\\
2455382.7267	&	0.0116	&	I	&	18.36	$\pm$	0.02	&	MDM 2.4-m	\\
2455382.7287	&	0.0132	&	I	&	18.35	$\pm$	0.02	&	MDM 2.4-m	\\
2455382.7307	&	0.0148	&	I	&	18.33	$\pm$	0.02	&	MDM 2.4-m	\\
2455382.7326	&	0.0164	&	I	&	18.33	$\pm$	0.02	&	MDM 2.4-m	\\
2455382.7346	&	0.0180	&	I	&	18.30	$\pm$	0.02	&	MDM 2.4-m	\\
2455382.7398	&	0.0222	&	I	&	18.28	$\pm$	0.02	&	MDM 2.4-m	\\
2455382.7418	&	0.0238	&	I	&	18.27	$\pm$	0.02	&	MDM 2.4-m	\\
2455382.7438	&	0.0254	&	I	&	18.25	$\pm$	0.02	&	MDM 2.4-m	\\
2455382.7457	&	0.0270	&	I	&	18.23	$\pm$	0.02	&	MDM 2.4-m	\\
2455382.7477	&	0.0286	&	I	&	18.25	$\pm$	0.02	&	MDM 2.4-m	\\
2455382.7497	&	0.0302	&	I	&	18.23	$\pm$	0.02	&	MDM 2.4-m	\\
2455382.7516	&	0.0318	&	I	&	18.21	$\pm$	0.02	&	MDM 2.4-m	\\
2455382.7536	&	0.0334	&	I	&	18.23	$\pm$	0.02	&	MDM 2.4-m	\\
2455382.7556	&	0.0350	&	I	&	18.20	$\pm$	0.02	&	MDM 2.4-m	\\
2455382.7576	&	0.0366	&	I	&	18.18	$\pm$	0.02	&	MDM 2.4-m	\\
2455382.7600	&	0.0386	&	I	&	18.18	$\pm$	0.02	&	MDM 2.4-m	\\
2455382.7620	&	0.0402	&	I	&	18.19	$\pm$	0.02	&	MDM 2.4-m	\\
2455382.7640	&	0.0419	&	I	&	18.17	$\pm$	0.02	&	MDM 2.4-m	\\
2455382.7659	&	0.0435	&	I	&	18.22	$\pm$	0.03	&	MDM 2.4-m	\\
2455382.7679	&	0.0451	&	I	&	18.17	$\pm$	0.03	&	MDM 2.4-m	\\
2455382.7699	&	0.0467	&	I	&	18.15	$\pm$	0.02	&	MDM 2.4-m	\\
2455382.7719	&	0.0483	&	I	&	18.07	$\pm$	0.02	&	MDM 2.4-m	\\
2455382.7738	&	0.0499	&	I	&	18.11	$\pm$	0.03	&	MDM 2.4-m	\\
2455382.7758	&	0.0515	&	I	&	18.14	$\pm$	0.02	&	MDM 2.4-m	\\
2455382.7778	&	0.0531	&	I	&	18.08	$\pm$	0.02	&	MDM 2.4-m	\\
2455382.7803	&	0.0551	&	I	&	18.04	$\pm$	0.02	&	MDM 2.4-m	\\
2455387.5582	&	0.9379	&	I	&	17.91	$\pm$	0.02	&	CTIO 0.9-m	\\
2455387.5625	&	0.9414	&	I	&	17.91	$\pm$	0.02	&	CTIO 0.9-m	\\
2455387.5668	&	0.9449	&	I	&	17.95	$\pm$	0.02	&	CTIO 0.9-m	\\
2455387.5711	&	0.9484	&	I	&	17.97	$\pm$	0.02	&	CTIO 0.9-m	\\
2455387.5754	&	0.9519	&	I	&	18.01	$\pm$	0.03	&	CTIO 0.9-m	\\
2455387.5797	&	0.9554	&	I	&	18.02	$\pm$	0.03	&	CTIO 0.9-m	\\
2455387.5840	&	0.9588	&	I	&	18.04	$\pm$	0.03	&	CTIO 0.9-m	\\
2455387.5883	&	0.9623	&	I	&	18.10	$\pm$	0.03	&	CTIO 0.9-m	\\
2455387.5926	&	0.9658	&	I	&	18.12	$\pm$	0.03	&	CTIO 0.9-m	\\
2455387.5969	&	0.9693	&	I	&	18.10	$\pm$	0.03	&	CTIO 0.9-m	\\
2455387.6015	&	0.9731	&	I	&	18.18	$\pm$	0.03	&	CTIO 0.9-m	\\
2455387.6058	&	0.9765	&	I	&	18.26	$\pm$	0.03	&	CTIO 0.9-m	\\
2455387.6101	&	0.9800	&	I	&	18.24	$\pm$	0.03	&	CTIO 0.9-m	\\
2455387.6144	&	0.9835	&	I	&	18.21	$\pm$	0.03	&	CTIO 0.9-m	\\
2455387.6187	&	0.9870	&	I	&	18.26	$\pm$	0.03	&	CTIO 0.9-m	\\
2455387.6230	&	0.9905	&	I	&	18.28	$\pm$	0.03	&	CTIO 0.9-m	\\
2455387.6273	&	0.9940	&	I	&	18.29	$\pm$	0.03	&	CTIO 0.9-m	\\
2455387.6316	&	0.9975	&	I	&	18.32	$\pm$	0.03	&	CTIO 0.9-m	\\
2455387.6359	&	0.0010	&	I	&	18.37	$\pm$	0.03	&	CTIO 0.9-m	\\
2455387.6402	&	0.0045	&	I	&	18.35	$\pm$	0.03	&	CTIO 0.9-m	\\
2455387.6445	&	0.0080	&	I	&	18.36	$\pm$	0.03	&	CTIO 0.9-m	\\
2455387.6488	&	0.0115	&	I	&	18.34	$\pm$	0.03	&	CTIO 0.9-m	\\
2455387.6531	&	0.0150	&	I	&	18.33	$\pm$	0.03	&	CTIO 0.9-m	\\
2455387.6574	&	0.0185	&	I	&	18.32	$\pm$	0.03	&	CTIO 0.9-m	\\
2455387.6617	&	0.0220	&	I	&	18.32	$\pm$	0.03	&	CTIO 0.9-m	\\
2455387.6660	&	0.0255	&	I	&	18.24	$\pm$	0.03	&	CTIO 0.9-m	\\
2455387.6703	&	0.0289	&	I	&	18.22	$\pm$	0.03	&	CTIO 0.9-m	\\
2455387.6746	&	0.0324	&	I	&	18.22	$\pm$	0.03	&	CTIO 0.9-m	\\
2455387.6789	&	0.0359	&	I	&	18.19	$\pm$	0.03	&	CTIO 0.9-m	\\
2455387.6832	&	0.0394	&	I	&	18.12	$\pm$	0.03	&	CTIO 0.9-m	\\
2455387.6875	&	0.0429	&	I	&	18.13	$\pm$	0.03	&	CTIO 0.9-m	\\
2455387.6918	&	0.0464	&	I	&	18.08	$\pm$	0.03	&	CTIO 0.9-m	\\
2455387.6961	&	0.0499	&	I	&	18.06	$\pm$	0.03	&	CTIO 0.9-m	\\
2455387.7004	&	0.0534	&	I	&	17.99	$\pm$	0.03	&	CTIO 0.9-m	\\
2455387.7047	&	0.0569	&	I	&	18.00	$\pm$	0.03	&	CTIO 0.9-m	\\
2455387.7090	&	0.0604	&	I	&	17.93	$\pm$	0.02	&	CTIO 0.9-m	\\
2455387.7133	&	0.0639	&	I	&	17.89	$\pm$	0.03	&	CTIO 0.9-m	\\
2455387.7176	&	0.0674	&	I	&	17.84	$\pm$	0.03	&	CTIO 0.9-m	\\
2455387.7219	&	0.0709	&	I	&	17.88	$\pm$	0.03	&	CTIO 0.9-m	\\
2455387.7262	&	0.0744	&	I	&	17.84	$\pm$	0.03	&	CTIO 0.9-m	\\
2455424.4697	&	0.9338	&	V	&	18.66	$\pm$	0.05	&	CHO 0.5-m	\\
2455424.4712	&	0.9351	&	V	&	18.59	$\pm$	0.04	&	CHO 0.5-m	\\
2455424.4727	&	0.9363	&	V	&	18.70	$\pm$	0.05	&	CHO 0.5-m	\\
2455424.4741	&	0.9375	&	V	&	18.71	$\pm$	0.05	&	CHO 0.5-m	\\
2455424.4756	&	0.9387	&	V	&	18.65	$\pm$	0.05	&	CHO 0.5-m	\\
2455424.4771	&	0.9399	&	V	&	18.62	$\pm$	0.05	&	CHO 0.5-m	\\
2455424.4786	&	0.9411	&	V	&	18.62	$\pm$	0.05	&	CHO 0.5-m	\\
2455424.4800	&	0.9423	&	V	&	18.58	$\pm$	0.05	&	CHO 0.5-m	\\
2455424.4877	&	0.9485	&	V	&	18.71	$\pm$	0.07	&	CHO 0.5-m	\\
2455424.4892	&	0.9497	&	V	&	18.70	$\pm$	0.06	&	CHO 0.5-m	\\
2455424.4907	&	0.9509	&	V	&	18.65	$\pm$	0.06	&	CHO 0.5-m	\\
2455424.4922	&	0.9521	&	V	&	18.68	$\pm$	0.06	&	CHO 0.5-m	\\
2455424.4937	&	0.9533	&	V	&	18.73	$\pm$	0.07	&	CHO 0.5-m	\\
2455424.4951	&	0.9545	&	V	&	18.69	$\pm$	0.08	&	CHO 0.5-m	\\
2455424.4966	&	0.9557	&	V	&	18.62	$\pm$	0.12	&	CHO 0.5-m	\\
2455424.4981	&	0.9569	&	V	&	19.12	$\pm$	0.10	&	CHO 0.5-m	\\
2455424.4995	&	0.9581	&	V	&	18.80	$\pm$	0.07	&	CHO 0.5-m	\\
2455424.5010	&	0.9593	&	V	&	18.76	$\pm$	0.07	&	CHO 0.5-m	\\
2455424.5025	&	0.9605	&	V	&	19.02	$\pm$	0.07	&	CHO 0.5-m	\\
2455424.5040	&	0.9617	&	V	&	18.76	$\pm$	0.05	&	CHO 0.5-m	\\
2455424.5054	&	0.9629	&	V	&	18.97	$\pm$	0.09	&	CHO 0.5-m	\\
2455424.5069	&	0.9641	&	V	&	18.83	$\pm$	0.07	&	CHO 0.5-m	\\
2455424.5084	&	0.9653	&	V	&	19.01	$\pm$	0.07	&	CHO 0.5-m	\\
2455424.5099	&	0.9665	&	V	&	18.84	$\pm$	0.06	&	CHO 0.5-m	\\
2455424.5113	&	0.9677	&	V	&	18.89	$\pm$	0.09	&	CHO 0.5-m	\\
2455424.5128	&	0.9689	&	V	&	18.92	$\pm$	0.06	&	CHO 0.5-m	\\
2455424.5143	&	0.9701	&	V	&	18.94	$\pm$	0.07	&	CHO 0.5-m	\\
2455424.5158	&	0.9713	&	V	&	19.03	$\pm$	0.06	&	CHO 0.5-m	\\
2455424.5172	&	0.9725	&	V	&	18.95	$\pm$	0.08	&	CHO 0.5-m	\\
2455424.5187	&	0.9737	&	V	&	19.03	$\pm$	0.06	&	CHO 0.5-m	\\
2455424.5202	&	0.9749	&	V	&	19.03	$\pm$	0.07	&	CHO 0.5-m	\\
2455424.5216	&	0.9761	&	V	&	19.01	$\pm$	0.06	&	CHO 0.5-m	\\
2455424.5231	&	0.9773	&	V	&	19.07	$\pm$	0.06	&	CHO 0.5-m	\\
2455424.5246	&	0.9785	&	V	&	18.92	$\pm$	0.06	&	CHO 0.5-m	\\
2455424.5261	&	0.9797	&	V	&	19.06	$\pm$	0.06	&	CHO 0.5-m	\\
2455424.5275	&	0.9809	&	V	&	19.05	$\pm$	0.06	&	CHO 0.5-m	\\
2455424.5290	&	0.9821	&	V	&	19.15	$\pm$	0.07	&	CHO 0.5-m	\\
2455424.5305	&	0.9833	&	V	&	19.19	$\pm$	0.07	&	CHO 0.5-m	\\
2455424.5320	&	0.9845	&	V	&	19.09	$\pm$	0.07	&	CHO 0.5-m	\\
2455424.5334	&	0.9857	&	V	&	19.19	$\pm$	0.07	&	CHO 0.5-m	\\
2455424.5349	&	0.9869	&	V	&	19.23	$\pm$	0.07	&	CHO 0.5-m	\\
2455424.5364	&	0.9881	&	V	&	19.19	$\pm$	0.07	&	CHO 0.5-m	\\
2455424.5379	&	0.9893	&	V	&	19.31	$\pm$	0.07	&	CHO 0.5-m	\\
2455424.5393	&	0.9905	&	V	&	19.18	$\pm$	0.07	&	CHO 0.5-m	\\
2455424.5408	&	0.9917	&	V	&	19.17	$\pm$	0.07	&	CHO 0.5-m	\\
2455424.5423	&	0.9929	&	V	&	19.24	$\pm$	0.08	&	CHO 0.5-m	\\
2455424.5438	&	0.9941	&	V	&	19.34	$\pm$	0.09	&	CHO 0.5-m	\\
2455424.5452	&	0.9953	&	V	&	19.20	$\pm$	0.07	&	CHO 0.5-m	\\
2455424.5467	&	0.9964	&	V	&	19.33	$\pm$	0.08	&	CHO 0.5-m	\\
2455424.5482	&	0.9976	&	V	&	19.25	$\pm$	0.08	&	CHO 0.5-m	\\
2455424.5496	&	0.9988	&	V	&	19.26	$\pm$	0.09	&	CHO 0.5-m	\\
2455424.5511	&	0.0000	&	V	&	19.32	$\pm$	0.09	&	CHO 0.5-m	\\
2455424.5526	&	0.0012	&	V	&	19.35	$\pm$	0.09	&	CHO 0.5-m	\\
2455424.5541	&	0.0024	&	V	&	19.71	$\pm$	0.15	&	CHO 0.5-m	\\
2455424.5555	&	0.0036	&	V	&	19.22	$\pm$	0.08	&	CHO 0.5-m	\\
2455424.5570	&	0.0048	&	V	&	19.30	$\pm$	0.08	&	CHO 0.5-m	\\
2455424.5585	&	0.0060	&	V	&	19.26	$\pm$	0.07	&	CHO 0.5-m	\\
2455424.5599	&	0.0072	&	V	&	19.39	$\pm$	0.09	&	CHO 0.5-m	\\
2455424.5614	&	0.0084	&	V	&	19.39	$\pm$	0.09	&	CHO 0.5-m	\\
2455424.5629	&	0.0096	&	V	&	19.25	$\pm$	0.08	&	CHO 0.5-m	\\
2455424.5644	&	0.0108	&	V	&	19.27	$\pm$	0.08	&	CHO 0.5-m	\\
2455424.5658	&	0.0120	&	V	&	19.31	$\pm$	0.08	&	CHO 0.5-m	\\
2455424.5673	&	0.0132	&	V	&	19.32	$\pm$	0.09	&	CHO 0.5-m	\\
2455424.5688	&	0.0144	&	V	&	19.07	$\pm$	0.07	&	CHO 0.5-m	\\
2455424.5703	&	0.0156	&	V	&	19.19	$\pm$	0.08	&	CHO 0.5-m	\\
2455424.5717	&	0.0168	&	V	&	19.20	$\pm$	0.08	&	CHO 0.5-m	\\
2455424.5732	&	0.0180	&	V	&	19.19	$\pm$	0.08	&	CHO 0.5-m	\\
2455424.5747	&	0.0192	&	V	&	19.26	$\pm$	0.09	&	CHO 0.5-m	\\
2455424.5762	&	0.0204	&	V	&	19.24	$\pm$	0.09	&	CHO 0.5-m	\\
2455424.5776	&	0.0216	&	V	&	19.14	$\pm$	0.08	&	CHO 0.5-m	\\
2455424.5791	&	0.0228	&	V	&	19.24	$\pm$	0.08	&	CHO 0.5-m	\\
2455424.5806	&	0.0240	&	V	&	19.26	$\pm$	0.11	&	CHO 0.5-m	\\
2455424.5821	&	0.0252	&	V	&	19.04	$\pm$	0.08	&	CHO 0.5-m	\\
2455424.5835	&	0.0264	&	V	&	19.04	$\pm$	0.07	&	CHO 0.5-m	\\
2455424.5850	&	0.0276	&	V	&	19.06	$\pm$	0.09	&	CHO 0.5-m	\\
2455424.5865	&	0.0288	&	V	&	19.30	$\pm$	0.10	&	CHO 0.5-m	\\
2455424.5880	&	0.0300	&	V	&	19.21	$\pm$	0.09	&	CHO 0.5-m	\\
2455424.5894	&	0.0312	&	V	&	18.87	$\pm$	0.07	&	CHO 0.5-m	\\
2455424.5909	&	0.0324	&	V	&	19.01	$\pm$	0.08	&	CHO 0.5-m	\\
2455424.5924	&	0.0336	&	V	&	19.05	$\pm$	0.08	&	CHO 0.5-m	\\
2455424.5939	&	0.0348	&	V	&	19.03	$\pm$	0.09	&	CHO 0.5-m	\\
2455424.5953	&	0.0360	&	V	&	19.09	$\pm$	0.11	&	CHO 0.5-m	\\
2455424.5968	&	0.0372	&	V	&	19.26	$\pm$	0.11	&	CHO 0.5-m	\\
2455424.5983	&	0.0384	&	V	&	19.16	$\pm$	0.10	&	CHO 0.5-m	\\
2455424.5998	&	0.0396	&	V	&	19.08	$\pm$	0.08	&	CHO 0.5-m	\\
2455424.6012	&	0.0408	&	V	&	18.86	$\pm$	0.07	&	CHO 0.5-m	\\
2455424.6027	&	0.0420	&	V	&	19.00	$\pm$	0.10	&	CHO 0.5-m	\\
2455424.6042	&	0.0432	&	V	&	18.80	$\pm$	0.08	&	CHO 0.5-m	\\
2455424.6056	&	0.0444	&	V	&	19.06	$\pm$	0.11	&	CHO 0.5-m	\\
2455424.6071	&	0.0455	&	V	&	19.07	$\pm$	0.10	&	CHO 0.5-m	\\
2455424.6086	&	0.0467	&	V	&	18.78	$\pm$	0.06	&	CHO 0.5-m	\\
2455424.6101	&	0.0479	&	V	&	18.85	$\pm$	0.07	&	CHO 0.5-m	\\
2455424.6115	&	0.0491	&	V	&	18.69	$\pm$	0.06	&	CHO 0.5-m	\\
2455424.6130	&	0.0503	&	V	&	18.64	$\pm$	0.06	&	CHO 0.5-m	\\
2455424.6145	&	0.0515	&	V	&	18.85	$\pm$	0.07	&	CHO 0.5-m	\\
2455424.6160	&	0.0527	&	V	&	18.74	$\pm$	0.07	&	CHO 0.5-m	\\
2455424.6174	&	0.0539	&	V	&	18.68	$\pm$	0.06	&	CHO 0.5-m	\\
2455424.6190	&	0.0552	&	V	&	18.79	$\pm$	0.07	&	CHO 0.5-m	\\
2455424.6205	&	0.0564	&	V	&	18.62	$\pm$	0.06	&	CHO 0.5-m	\\
2455424.6219	&	0.0576	&	V	&	18.65	$\pm$	0.06	&	CHO 0.5-m	\\
2455424.6234	&	0.0588	&	V	&	18.62	$\pm$	0.06	&	CHO 0.5-m	\\
2455424.6249	&	0.0600	&	V	&	18.74	$\pm$	0.07	&	CHO 0.5-m	\\
2455424.6263	&	0.0612	&	V	&	18.72	$\pm$	0.07	&	CHO 0.5-m	\\
2455424.6278	&	0.0624	&	V	&	18.69	$\pm$	0.08	&	CHO 0.5-m	\\
2455424.6293	&	0.0636	&	V	&	18.54	$\pm$	0.06	&	CHO 0.5-m	\\
2455424.6308	&	0.0648	&	V	&	18.53	$\pm$	0.07	&	CHO 0.5-m	\\
2455424.6323	&	0.0660	&	V	&	18.55	$\pm$	0.06	&	CHO 0.5-m	\\
2455424.6337	&	0.0672	&	V	&	18.52	$\pm$	0.07	&	CHO 0.5-m	\\
2455424.6352	&	0.0684	&	V	&	18.52	$\pm$	0.06	&	CHO 0.5-m	\\
2455424.6367	&	0.0696	&	V	&	18.50	$\pm$	0.06	&	CHO 0.5-m	\\
2455424.6381	&	0.0708	&	V	&	18.50	$\pm$	0.06	&	CHO 0.5-m	\\
2455424.6396	&	0.0720	&	V	&	18.44	$\pm$	0.05	&	CHO 0.5-m	\\
2455424.6411	&	0.0732	&	V	&	18.43	$\pm$	0.05	&	CHO 0.5-m	\\
2455424.6426	&	0.0744	&	V	&	18.37	$\pm$	0.05	&	CHO 0.5-m	\\
2455424.6440	&	0.0755	&	V	&	18.41	$\pm$	0.05	&	CHO 0.5-m	\\
2455424.6455	&	0.0768	&	V	&	18.40	$\pm$	0.06	&	CHO 0.5-m	\\
2455424.6470	&	0.0779	&	V	&	18.39	$\pm$	0.06	&	CHO 0.5-m	\\
2455424.6485	&	0.0791	&	V	&	18.46	$\pm$	0.06	&	CHO 0.5-m	\\
2455424.6499	&	0.0803	&	V	&	18.35	$\pm$	0.05	&	CHO 0.5-m	\\
2455424.6514	&	0.0815	&	V	&	18.43	$\pm$	0.06	&	CHO 0.5-m	\\
2455424.6529	&	0.0827	&	V	&	18.42	$\pm$	0.06	&	CHO 0.5-m	\\
2455424.6543	&	0.0839	&	V	&	18.51	$\pm$	0.06	&	CHO 0.5-m	\\
2455424.6558	&	0.0851	&	V	&	18.30	$\pm$	0.05	&	CHO 0.5-m	\\
2455424.6573	&	0.0863	&	V	&	18.34	$\pm$	0.05	&	CHO 0.5-m	\\
2455424.6588	&	0.0876	&	V	&	18.33	$\pm$	0.05	&	CHO 0.5-m	\\
2455424.6603	&	0.0888	&	V	&	18.40	$\pm$	0.05	&	CHO 0.5-m	\\
2455424.6618	&	0.0900	&	V	&	18.40	$\pm$	0.06	&	CHO 0.5-m	\\
2455424.6633	&	0.0912	&	V	&	18.40	$\pm$	0.05	&	CHO 0.5-m	\\
2455424.6647	&	0.0924	&	V	&	18.33	$\pm$	0.05	&	CHO 0.5-m	\\
2455424.6662	&	0.0936	&	V	&	18.37	$\pm$	0.07	&	CHO 0.5-m	\\
2455424.6677	&	0.0948	&	V	&	18.24	$\pm$	0.05	&	CHO 0.5-m	\\
2455424.6692	&	0.0960	&	V	&	18.27	$\pm$	0.06	&	CHO 0.5-m	\\
2455424.6706	&	0.0972	&	V	&	18.25	$\pm$	0.05	&	CHO 0.5-m	\\
2455424.6721	&	0.0984	&	V	&	18.31	$\pm$	0.06	&	CHO 0.5-m	\\
2455424.6736	&	0.0995	&	V	&	18.34	$\pm$	0.05	&	CHO 0.5-m	\\
2455424.6751	&	0.1008	&	V	&	18.32	$\pm$	0.05	&	CHO 0.5-m	\\
2455424.6766	&	0.1020	&	V	&	18.43	$\pm$	0.06	&	CHO 0.5-m	\\
2455424.6781	&	0.1032	&	V	&	18.29	$\pm$	0.05	&	CHO 0.5-m	\\
2455424.6795	&	0.1044	&	V	&	18.38	$\pm$	0.06	&	CHO 0.5-m	\\
2455424.6811	&	0.1057	&	V	&	18.15	$\pm$	0.05	&	CHO 0.5-m	\\
2455424.6826	&	0.1069	&	V	&	18.22	$\pm$	0.04	&	CHO 0.5-m	\\
2455424.6840	&	0.1081	&	V	&	18.15	$\pm$	0.05	&	CHO 0.5-m	\\
2455424.6855	&	0.1092	&	V	&	18.17	$\pm$	0.05	&	CHO 0.5-m	\\
2455424.6870	&	0.1105	&	V	&	18.20	$\pm$	0.05	&	CHO 0.5-m	\\
2455424.6885	&	0.1117	&	V	&	18.24	$\pm$	0.05	&	CHO 0.5-m	\\
2455424.6900	&	0.1129	&	V	&	18.30	$\pm$	0.05	&	CHO 0.5-m	\\
2455424.6914	&	0.1141	&	V	&	18.19	$\pm$	0.06	&	CHO 0.5-m	\\
2455424.6931	&	0.1154	&	V	&	18.16	$\pm$	0.05	&	CHO 0.5-m	\\
2455424.6946	&	0.1167	&	V	&	18.13	$\pm$	0.07	&	CHO 0.5-m	\\
2455424.6961	&	0.1179	&	V	&	18.00	$\pm$	0.06	&	CHO 0.5-m	\\
2455424.6977	&	0.1191	&	V	&	18.04	$\pm$	0.07	&	CHO 0.5-m	\\
2455424.6992	&	0.1204	&	V	&	18.14	$\pm$	0.10	&	CHO 0.5-m	\\
2455424.7007	&	0.1216	&	V	&	18.06	$\pm$	0.08	&	CHO 0.5-m	\\
2455424.7023	&	0.1229	&	V	&	18.13	$\pm$	0.08	&	CHO 0.5-m	\\
2455424.7039	&	0.1242	&	V	&	17.98	$\pm$	0.09	&	CHO 0.5-m	\\
2455424.7056	&	0.1256	&	V	&	17.90	$\pm$	0.10	&	CHO 0.5-m	\\
2455424.7072	&	0.1268	&	V	&	17.99	$\pm$	0.12	&	CHO 0.5-m	\\
2455424.7102	&	0.1294	&	V	&	17.89	$\pm$	0.13	&	CHO 0.5-m	\\
2455424.7132	&	0.1318	&	V	&	17.94	$\pm$	0.14	&	CHO 0.5-m	\\

\enddata
\end{deluxetable}

\begin{deluxetable}{llll}
\tabletypesize{\scriptsize}
\tablecaption{Trend Line for Eruption Light Curve
\label{tbl5}}
\tablewidth{0pt}
\tablehead{
\colhead{HJD}   &
\colhead{$T-T_0$ (days)}   &
\colhead{$V_{trend}$ (mag)}   &
\colhead{Comments}
}
\startdata

2455224.32	&	0.00	&	18.00	&	Eruption start ($T_0$)	\\
2455224.69	&	0.37	&	7.60	&	Peak	\\
2455226	&	1.68	&	9.40	&	$t_2=1.7$ days	\\
2455228	&	3.68	&	10.43	&	$t_3=3.6$ days	\\
2455230	&	5.68	&	11.45	&	\ldots	\\
2455231	&	6.68	&	11.70	&	\ldots	\\
2455232	&	7.68	&	12.35	&	\ldots	\\
2455233	&	8.68	&	13.00	&	Start of early flares	\\
2455234	&	9.68	&	13.15	&	Short  0.5 mag flares	\\
2455236	&	11.68	&	13.90	&	Short  0.5 mag flares	\\
2455238	&	13.68	&	14.00	&	Onset of plateau, SSS, eclipses	\\
2455240	&	15.68	&	14.05	&	\ldots	\\
2455242	&	17.68	&	14.15	&	\ldots	\\
2455244	&	19.68	&	14.20	&	\ldots	\\
2455246	&	21.68	&	14.20	&	\ldots	\\
2455248	&	23.68	&	14.30	&	Onset of flickering	\\
2455250	&	25.68	&	14.40	&	Onset of secondary eclipses	\\
2455252	&	27.68	&	14.50	&	\ldots	\\
2455254	&	29.68	&	14.70	&	\ldots	\\
2455255	&	30.68	&	14.90	&	\ldots	\\
2455256	&	31.68	&	15.00	&	End of plateau, sec. eclipses	\\
2455258	&	33.68	&	15.30	&	\ldots	\\
2455259	&	34.68	&	15.50	&	\ldots	\\
2455260	&	35.68	&	15.70	&	\ldots	\\
2455261	&	36.68	&	15.75	&	\ldots	\\
2455262	&	37.68	&	16.00	&	\ldots	\\
2455263	&	38.68	&	16.10	&	\ldots	\\
2455264	&	39.68	&	16.40	&	\ldots	\\
2455265	&	40.68	&	16.60	&	\ldots	\\
2455266	&	41.68	&	16.90	&	Onset of optical dips, plateau	\\
2455267	&	42.68	&	16.80	&	\ldots	\\
2455268	&	43.68	&	16.70	&	\ldots	\\
2455269	&	44.68	&	16.90	&	\ldots	\\
2455271	&	46.68	&	16.90	&	\ldots	\\
2455272	&	47.68	&	17.20	&	\ldots	\\
2455273	&	48.68	&	16.80	&	\ldots	\\
2455274	&	49.68	&	16.70	&	\ldots	\\
2455275	&	50.68	&	16.80	&	\ldots	\\
2455276	&	51.68	&	16.90	&	\ldots	\\
2455278	&	53.68	&	16.90	&	End of second plateau	\\
2455279	&	54.68	&	17.00	&	\ldots	\\
2455280	&	55.68	&	16.60	&	\ldots	\\
2455281	&	56.68	&	16.80	&	\ldots	\\
2455283	&	58.68	&	17.45	&	\ldots	\\
2455285	&	60.68	&	17.30	&	End of optical dips	\\
2455287	&	62.68	&	17.20	&	\ldots	\\
2455288	&	63.68	&	18.10	&	\ldots	\\
2455289	&	64.68	&	17.60	&	\ldots	\\
2455291	&	66.68	&	18.00	&	End of eruption	\\
2455301	&	76.68	&	18.00	&	\ldots	\\

\enddata
\end{deluxetable}

\begin{deluxetable}{lllllll}
\tabletypesize{\scriptsize}
\tablecaption{$V-V_{trend}$ Phased Light Curve Templates
\label{tbl6}}
\tablewidth{0pt}
\tablehead{
\colhead{Phase}   &
\colhead{Days 15-21}   &
\colhead{Days 21-26}   &
\colhead{Days 26-32}   &
\colhead{Days 32-41}   &
\colhead{Days 41-54}   &
\colhead{Days 54-67}
}
\startdata

0.00	&	0.60	&	0.75	&	1.16	&	1.36	&	0.95	&	1.11	\\
0.01	&	0.60	&	0.75	&	1.16	&	1.36	&	0.95	&	1.11	\\
0.02	&	0.60	&	0.72	&	1.09	&	1.30	&	0.92	&	1.11	\\
0.03	&	0.58	&	0.66	&	0.99	&	1.17	&	0.82	&	1.05	\\
0.04	&	0.55	&	0.59	&	0.88	&	1.05	&	0.75	&	0.90	\\
0.05	&	0.48	&	0.52	&	0.73	&	0.91	&	0.66	&	0.70	\\
0.06	&	0.41	&	0.48	&	0.66	&	0.70	&	0.54	&	0.54	\\
0.07	&	0.35	&	0.35	&	0.61	&	0.55	&	0.40	&	0.40	\\
0.08	&	0.32	&	0.28	&	0.55	&	0.40	&	0.30	&	0.33	\\
0.09	&	0.26	&	0.20	&	0.49	&	0.30	&	0.23	&	0.28	\\
0.10	&	0.22	&	0.16	&	0.40	&	0.25	&	0.16	&	0.22	\\
0.11	&	0.18	&	0.10	&	0.35	&	0.20	&	0.10	&	0.17	\\
0.12	&	0.15	&	0.03	&	0.30	&	0.16	&	0.02	&	0.12	\\
0.13	&	0.12	&	0.00	&	0.25	&	0.12	&	0.00	&	0.09	\\
0.14	&	0.09	&	0.03	&	0.20	&	0.08	&	0.00	&	0.05	\\
0.15	&	0.06	&	0.05	&	0.15	&	0.05	&	0.00	&	0.02	\\
0.16	&	0.03	&	0.06	&	0.11	&	0.02	&	0.00	&	0.00	\\
0.17	&	0.02	&	0.07	&	0.07	&	0.00	&	0.00	&	0.00	\\
0.18	&	0.01	&	0.08	&	0.04	&	-0.02	&	0.00	&	0.00	\\
0.20	&	0.00	&	0.08	&	0.01	&	-0.04	&	0.00	&	0.00	\\
0.24	&	0.00	&	0.06	&	0.01	&	-0.02	&	0.00	&	0.00	\\
0.28	&	0.00	&	0.01	&	0.01	&	0.03	&	0.00	&	0.00	\\
0.32	&	0.00	&	0.00	&	0.03	&	0.07	&	0.00	&	0.00	\\
0.36	&	0.10	&	0.03	&	0.06	&	0.04	&	0.00	&	0.00	\\
0.40	&	0.18	&	0.08	&	0.12	&	0.01	&	0.00	&	0.00	\\
0.44	&	0.18	&	0.14	&	0.28	&	0.05	&	0.00	&	0.00	\\
0.48	&	0.18	&	0.20	&	0.28	&	0.09	&	0.00	&	0.00	\\
0.50	&	0.18	&	0.20	&	0.28	&	0.10	&	0.00	&	0.00	\\
0.52	&	0.18	&	0.20	&	0.28	&	0.10	&	0.00	&	0.00	\\
0.56	&	0.18	&	0.20	&	0.28	&	0.08	&	0.00	&	0.00	\\
0.60	&	0.16	&	0.16	&	0.23	&	0.04	&	0.00	&	0.00	\\
0.64	&	0.12	&	0.08	&	0.20	&	0.09	&	0.00	&	0.00	\\
0.68	&	0.09	&	0.07	&	0.14	&	0.19	&	0.00	&	0.00	\\
0.72	&	0.12	&	0.09	&	0.12	&	0.25	&	0.00	&	0.00	\\
0.76	&	0.15	&	0.15	&	0.14	&	0.27	&	0.00	&	0.00	\\
0.80	&	0.18	&	0.18	&	0.21	&	0.28	&	0.00	&	0.00	\\
0.82	&	0.20	&	0.20	&	0.28	&	0.30	&	0.00	&	0.00	\\
0.83	&	0.22	&	0.21	&	0.30	&	0.33	&	0.00	&	0.00	\\
0.84	&	0.23	&	0.22	&	0.32	&	0.37	&	0.00	&	0.01	\\
0.85	&	0.24	&	0.23	&	0.34	&	0.41	&	0.00	&	0.03	\\
0.86	&	0.26	&	0.24	&	0.37	&	0.46	&	0.02	&	0.05	\\
0.87	&	0.28	&	0.25	&	0.40	&	0.53	&	0.03	&	0.07	\\
0.88	&	0.30	&	0.27	&	0.45	&	0.60	&	0.04	&	0.10	\\
0.89	&	0.32	&	0.30	&	0.52	&	0.65	&	0.06	&	0.15	\\
0.90	&	0.34	&	0.33	&	0.59	&	0.69	&	0.09	&	0.27	\\
0.91	&	0.37	&	0.37	&	0.66	&	0.73	&	0.12	&	0.31	\\
0.92	&	0.40	&	0.42	&	0.72	&	0.79	&	0.20	&	0.33	\\
0.93	&	0.44	&	0.49	&	0.79	&	0.90	&	0.31	&	0.36	\\
0.94	&	0.47	&	0.57	&	0.88	&	0.97	&	0.42	&	0.50	\\
0.95	&	0.51	&	0.62	&	0.97	&	1.02	&	0.53	&	0.60	\\
0.96	&	0.55	&	0.68	&	1.04	&	1.08	&	0.64	&	0.70	\\
0.97	&	0.60	&	0.73	&	1.10	&	1.15	&	0.77	&	0.80	\\
0.98	&	0.60	&	0.75	&	1.13	&	1.26	&	0.86	&	0.95	\\
0.99	&	0.60	&	0.75	&	1.16	&	1.36	&	0.95	&	1.11	\\
1.00	&	0.60	&	0.75	&	1.16	&	1.36	&	0.95	&	1.11	\\

\enddata
\end{deluxetable}

\begin{deluxetable}{llllllll}
\tabletypesize{\scriptsize}
\tablecaption{Eclipses During Eruptions
\label{tbl7}}
\tablewidth{0pt}
\tablehead{
\colhead{UT Date}   &
\colhead{$T-T_0$ (days)}   &
\colhead{Observer}   &
\colhead{HJD minimum}   &
\colhead{Amp. (mag)}   &
\colhead{FWHM (days)}   &
\colhead{$N$}   &
\colhead{O-C (days)}
}
\startdata

1999 Mar 16	&	19.2	&	Matsumoto\tablenotemark{a}	&	2451254.2110	$\pm$	0.0100		&	\ldots		&	\ldots	&	16	&	-0.0165	\\
1999 Mar 27	&	30.3	&	Ouda\tablenotemark{b}	&	2451265.3060	$\pm$	0.0100		&	\ldots		&	\ldots	&	25	&	0.0036	\\
1999 Apr 17	&	51.2	&	Thoroughgood\tablenotemark{c}	&	2451286.2143	$\pm$	0.0050		&	\ldots		&	\ldots	&	42	&	-0.0074	\\
2010 Feb 12	&	15.6	&	Stein	&	2455239.9600	$\pm$	0.0200		&	0.62		&	\ldots	&	3255	&	-0.0090	\\
2010 Feb 17	&	20.6	&	Oksanen\tablenotemark{d}	&	2455244.8778	$\pm$	0.0005	\tablenotemark{e}	&	0.70		&	0.248	&	3259	&	-0.0134	\\
2010 Feb 19	&	23.0	&	Tan	&	2455247.3505	$\pm$	0.0018		&	0.79		&	\ldots	&	3261	&	-0.0018	\\
2010 Feb 22	&	25.5	&	Oksanen	&	2455249.8047	$\pm$	0.0008		&	0.77		&	\ldots	&	3263	&	-0.0087	\\
2010 Feb 24	&	27.9	&	Tan, Stockdale	&	2455252.2681	$\pm$	0.0013		&	1.18		&	0.205	&	3265	&	-0.0064	\\
2010 Mar 5	&	36.6	&	Oksanen	&	2455260.8838	$\pm$	0.0010		&	1.35		&	0.204	&	3272	&	-0.0045	\\
2010 Mar 10	&	41.5	&	Oksanen	&	2455265.8097	$\pm$	0.0015		&	1.02		&	0.150	&	3276	&	-0.0008	\\
2010 Mar 12	&	43.9	&	Stockdale	&	2455268.2625	$\pm$	0.0020		&	1.20		&	\ldots	&	3278	&	-0.0091	\\
2010 Mar 15	&	46.4	&	Oksanen	&	2455270.7446	$\pm$	0.0009		&	0.87		&	\ldots	&	3280	&	0.0119	\\
2010 Mar 16	&	47.6	&	Krajci	&	2455271.9637	$\pm$	0.0031		&	0.92		&	\ldots	&	3281	&	0.0005	\\
2010 Mar 26	&	57.5	&	Oksanen	&	2455281.8158	$\pm$	0.0012		&	1.19		&	0.135	&	3289	&	0.0082	\\
2010 Mar 31	&	62.4	&	Oksanen	&	2455286.7411	$\pm$	0.0025		&	1.14		&	0.135	&	3293	&	0.0113	\\
2010 May 18	&	110.4	&	Schaefer	&	2455334.7211	$\pm$	0.0009		&	\ldots		&	\ldots	&	3332	&	0.0000	\\
2010 Jun 29	&	152.2	&	Schaefer	&	2455376.5650	$\pm$	0.0035		&	0.80	\tablenotemark{f}	&	\ldots	&	3366	&	0.0053	\\
2010 Jul 5	&	158.4	&	Lepine	&	2455382.7126	$\pm$	0.0008		&	0.98	\tablenotemark{f}	&	\ldots	&	3371	&	0.0001	\\
2010 Jul 10	&	163.3	&	Schaefer	&	2455387.6395	$\pm$	0.0010		&	0.91	\tablenotemark{f}	&	\ldots	&	3375	&	0.0048	\\
2010 Aug 16	&	200.2	&	Oksanen	&	2455424.5565	$\pm$	0.0010		&	0.95		&	0.103	&	3405	&	0.0054	\\

\enddata
\tablenotetext{a}{Matsumoto et al. 2003}
\tablenotetext{b}{Data from VSNET; http://www.kusastro.kyoto-u.ac.jp/vsnet/Novae/usco.html}
\tablenotetext{c}{Thoroughgood et al. 2001}
\tablenotetext{d}{Oksanen covers the ingress and minimum, while Stein, Harris, Krajci, and Henden cover the egress}
\tablenotetext{e}{The formal uncertainty is $\pm$0.0005 days, but possible uncertainties in offset the offset from Oksanen's photometry to the other observers might make the time of minimum uncertain up to $\pm$0.01 days}
\tablenotetext{f}{Amplitude in I-band, not V-band}
\end{deluxetable}

\clearpage
\begin{figure}
\epsscale{1.0}
\plotone{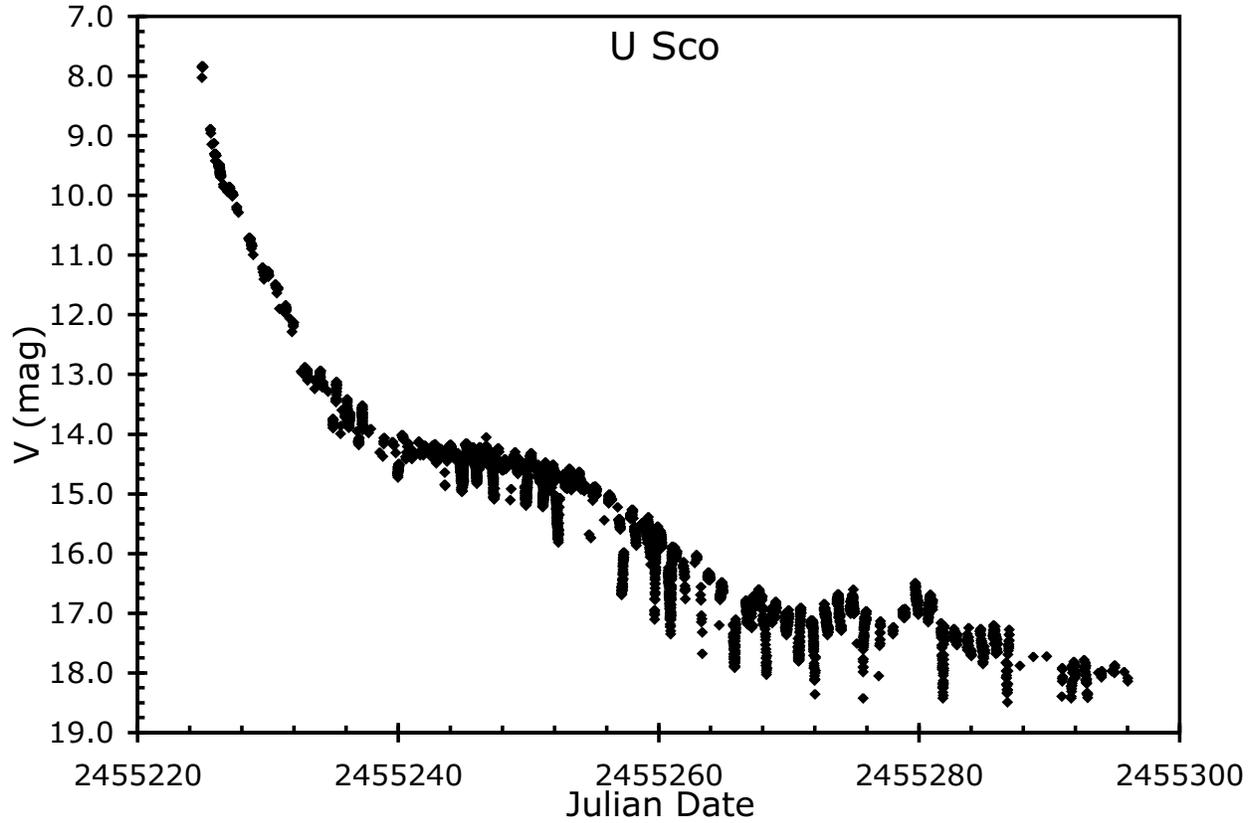}
\caption{
U Sco eruption light curve.  This light curve is based on 16,995 binned V-band magnitudes.  With a total of 36,776 magnitudes and complete coverage throughout the entire 67 days of the eruption, we cover the whole U Sco eruption with an average rate of one magnitude every 2.6 minutes.  We see the very fast decline, the startup of eclipses when the first plateau begins, and a second late plateau just above the quiescent level.}
\end{figure}

\clearpage
\begin{figure}
\epsscale{1.0}
\plotone{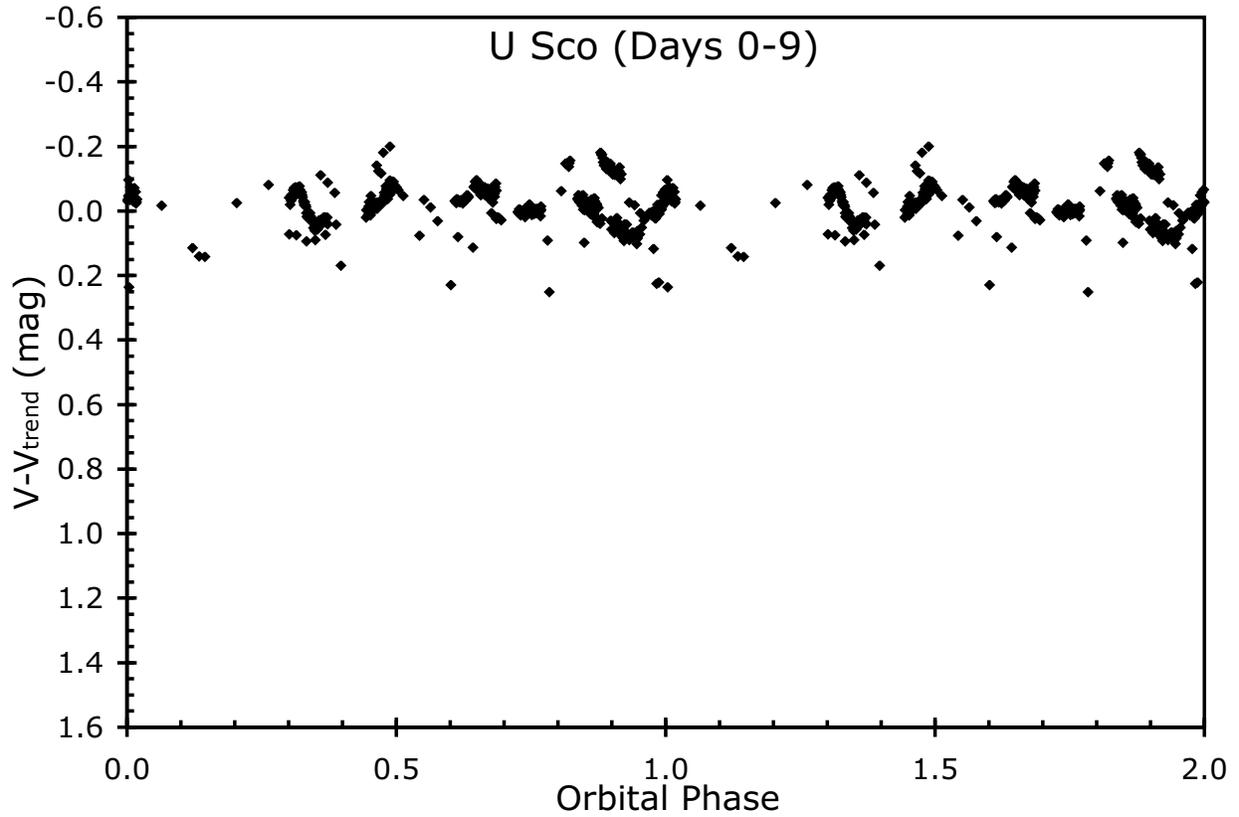}
\caption{
Detrended phased light curve for days 0-9.  Figures 2-9 show the $V-V_{trend}$ magnitudes plotted (with a doubling of phase) for a series of time intervals throughout the eruption.  For this figure (covering days 0-9), the initial fast decline is relatively smooth, so the detrended light curve appears flat with no significant variations with orbital phase.}
\end{figure}

\clearpage
\begin{figure}
\epsscale{1.0}
\plotone{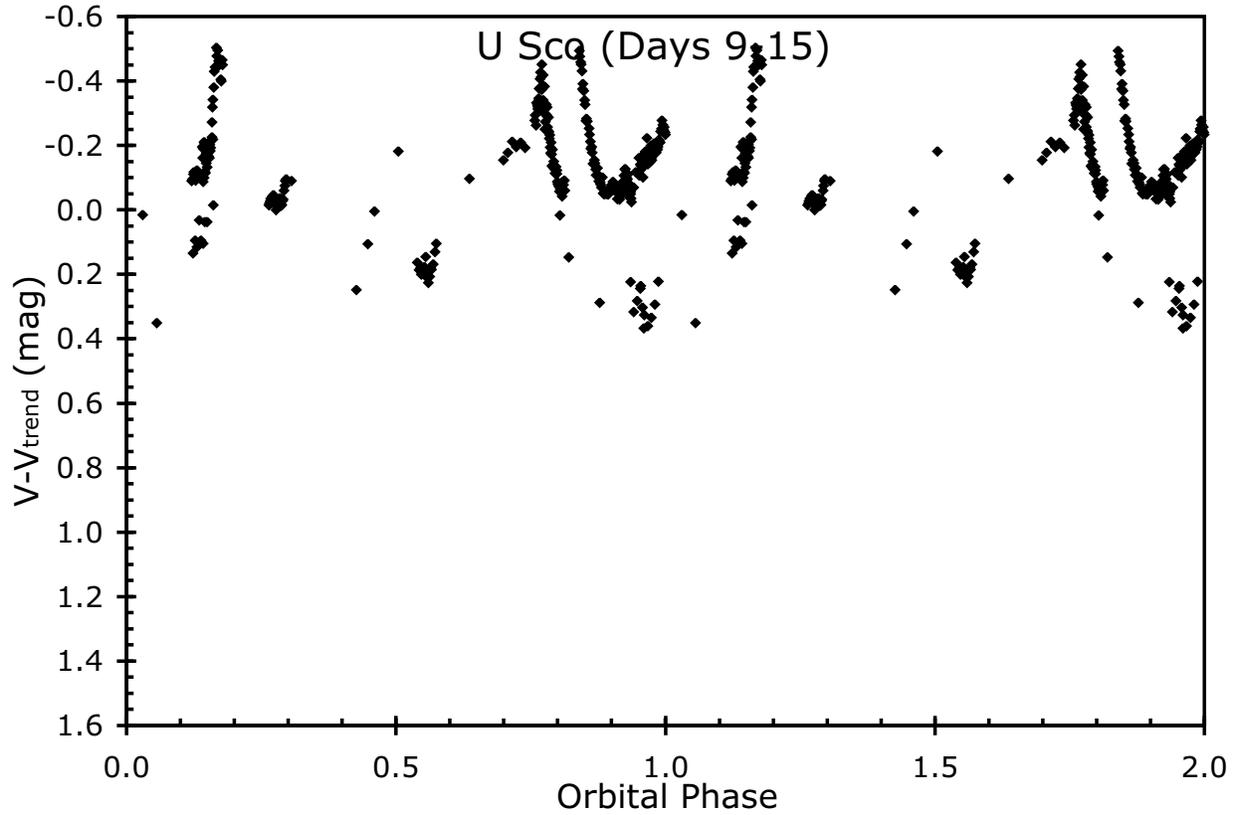}
\caption{
Days 9-15.  We see random flares with amplitude of half a magnitude and durations of half an hour.  These flares can only come from small regions of the shell which suddenly brighten with a luminosity rivaling that of the entire shell.  The cause of these flares is currently unknown.}
\end{figure}

\clearpage
\begin{figure}
\epsscale{1.0}
\plotone{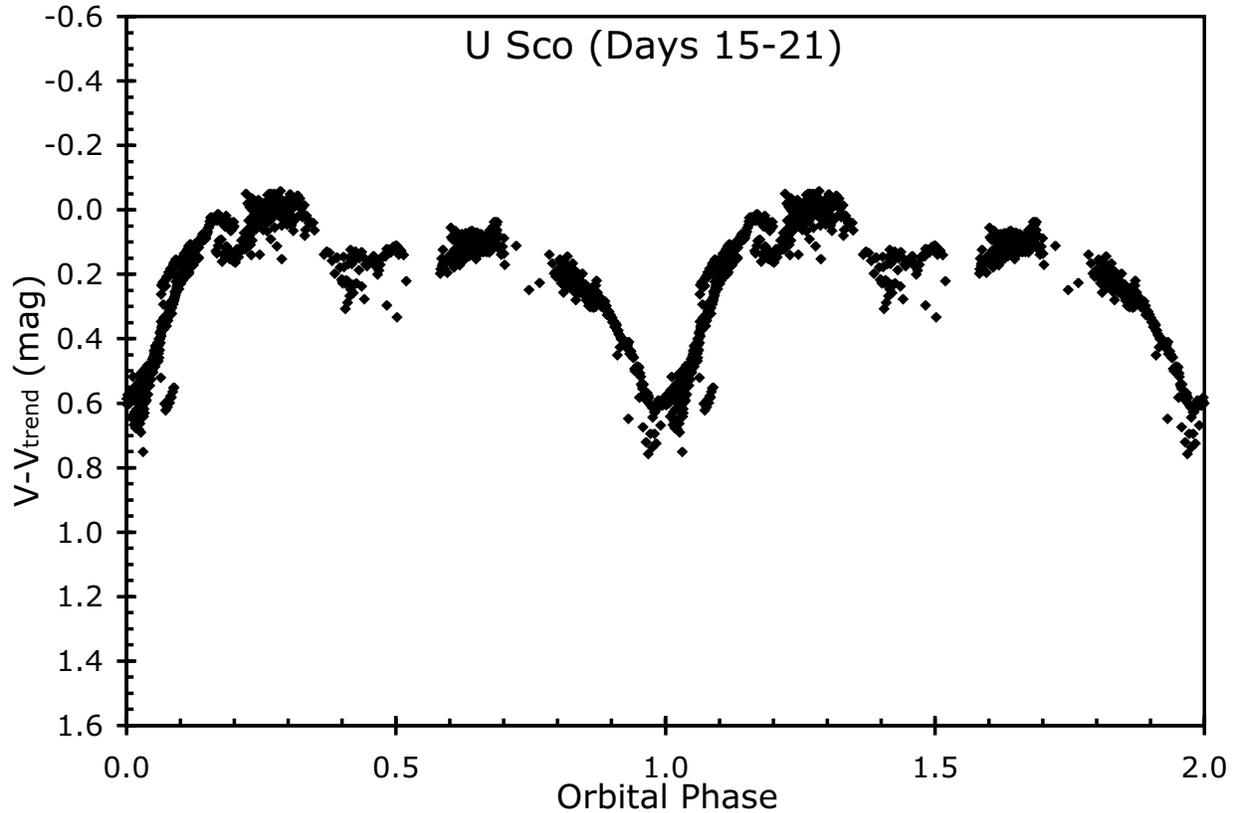}
\caption{
Days 15-21.  The eclipses suddenly start up sometime between days 12.0 and 15.6.  Coincident with this is the sudden turn-on of the supersoft X-ray source (SSS), and the start of the plateau.  All three phenomena are explained by the outer shell thinning enough so that the inner binary system becomes visible.  Then, the soft X-ray photons from near the surface of the white dwarf can escape, the nearly constant light from the binary provides the steady light making the plateau in the overall light curve, and the eclipses can be seen. }
\end{figure}

\clearpage
\begin{figure}
\epsscale{1.0}
\plotone{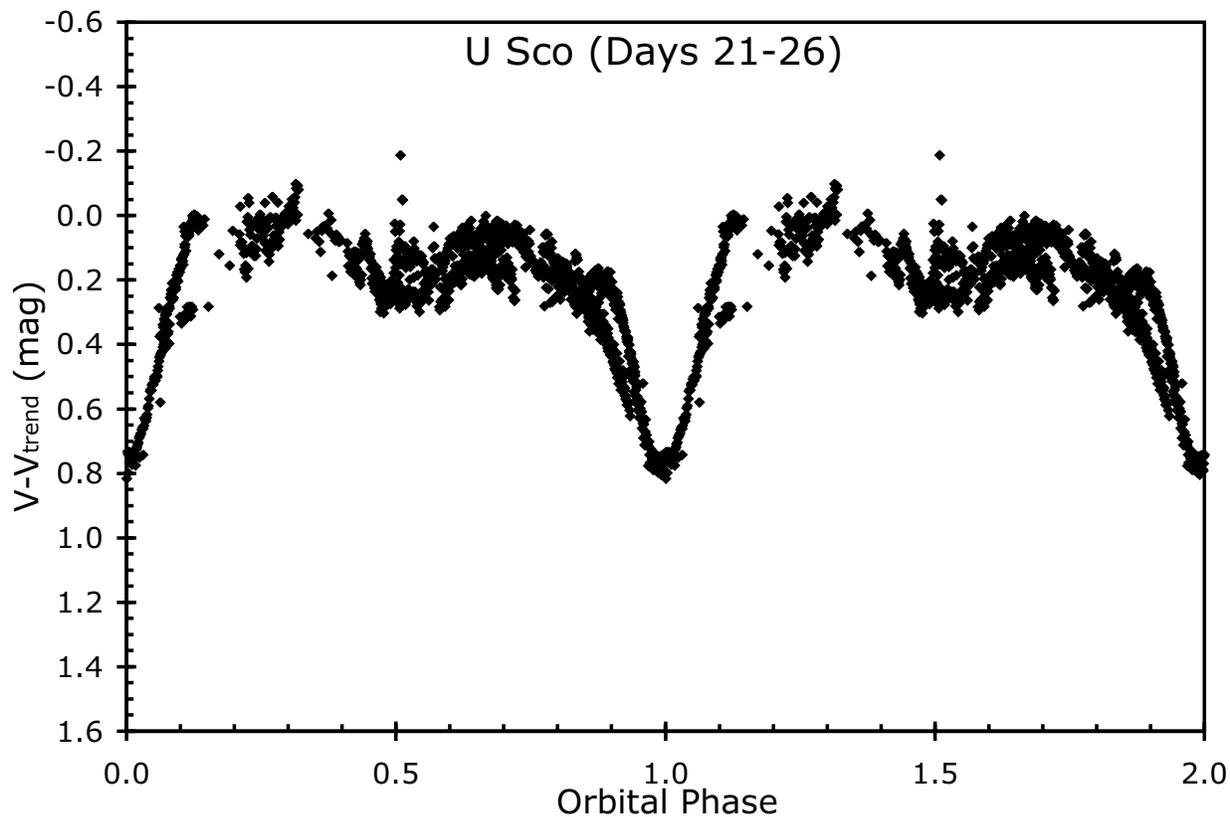}
\caption{
Days 21-26.  The eclipses deepen and become slightly shorter in duration in comparison with the previous week.  From days 15-26, the light curve shows a curious asymmetry between the elongations at phase 0.25 and 0.75.  This asymmetry could be caused by the illumination of the inner side of the accretion stream ahead of the companion star.  Eclipse mapping shows that all of the optical light is configured as an apparently uniform sphere with radius 4.1 R$_{\odot}$, which can only be the emission from the usual nova wind being driven off the white dwarf by the continuing nuclear burning near its surface.}
\end{figure}

\clearpage
\begin{figure}
\epsscale{1.0}
\plotone{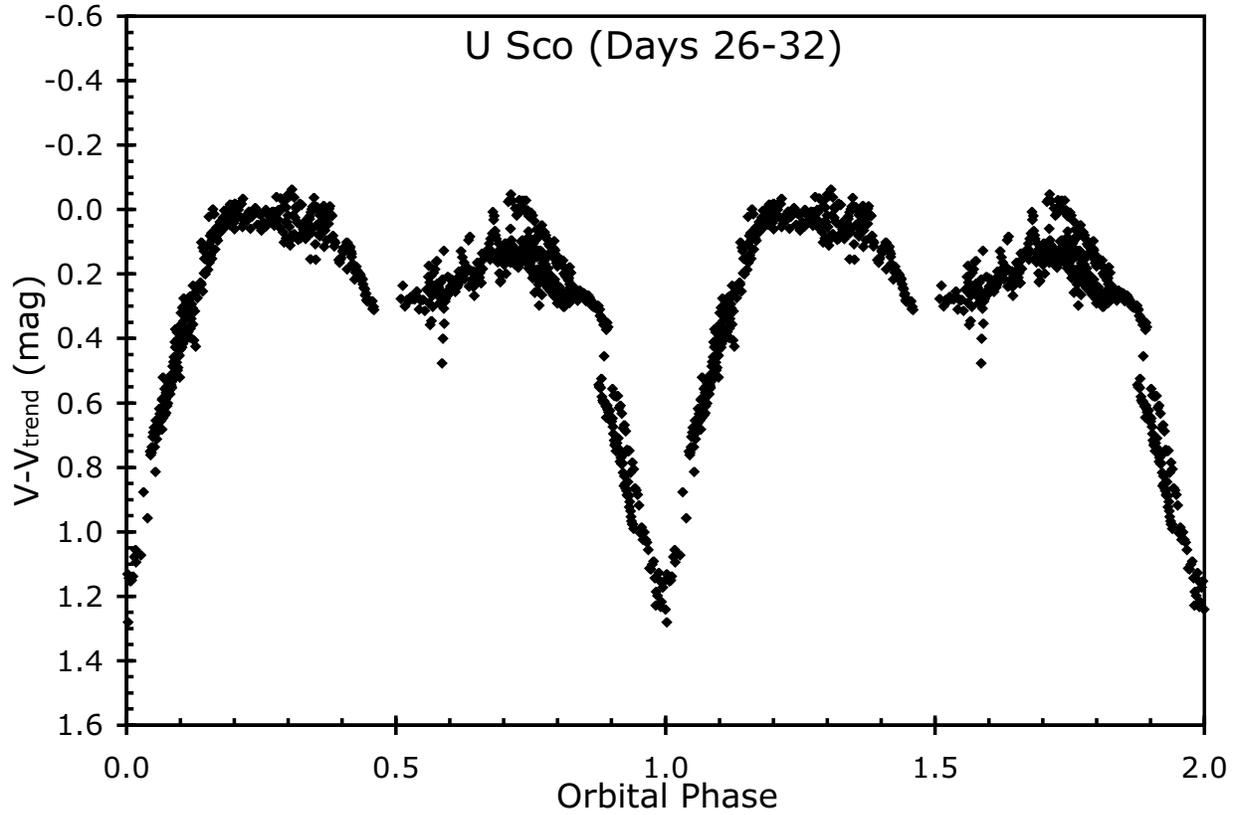}
\caption{
Days 26-32.  The eclipses become even deeper and broader, and the secondary eclipse is apparent.  The asymmetry between phases 0.25 and 0.75 has become less prominent.  Eclipse mapping shows that the configuration of the optical light source has changed completely, with there now being no light coming from any location except the orbital plane, so the wind is no longer contributing much optical light, but rather the optical light is coming from a large optical disk with radius roughly 3.4 R$_{\odot}$ that is faint in the center.  This shows that the accretion disk has been re-established but has not yet had time to work material into its central region.}
\end{figure}

\clearpage
\begin{figure}
\epsscale{1.0}
\plotone{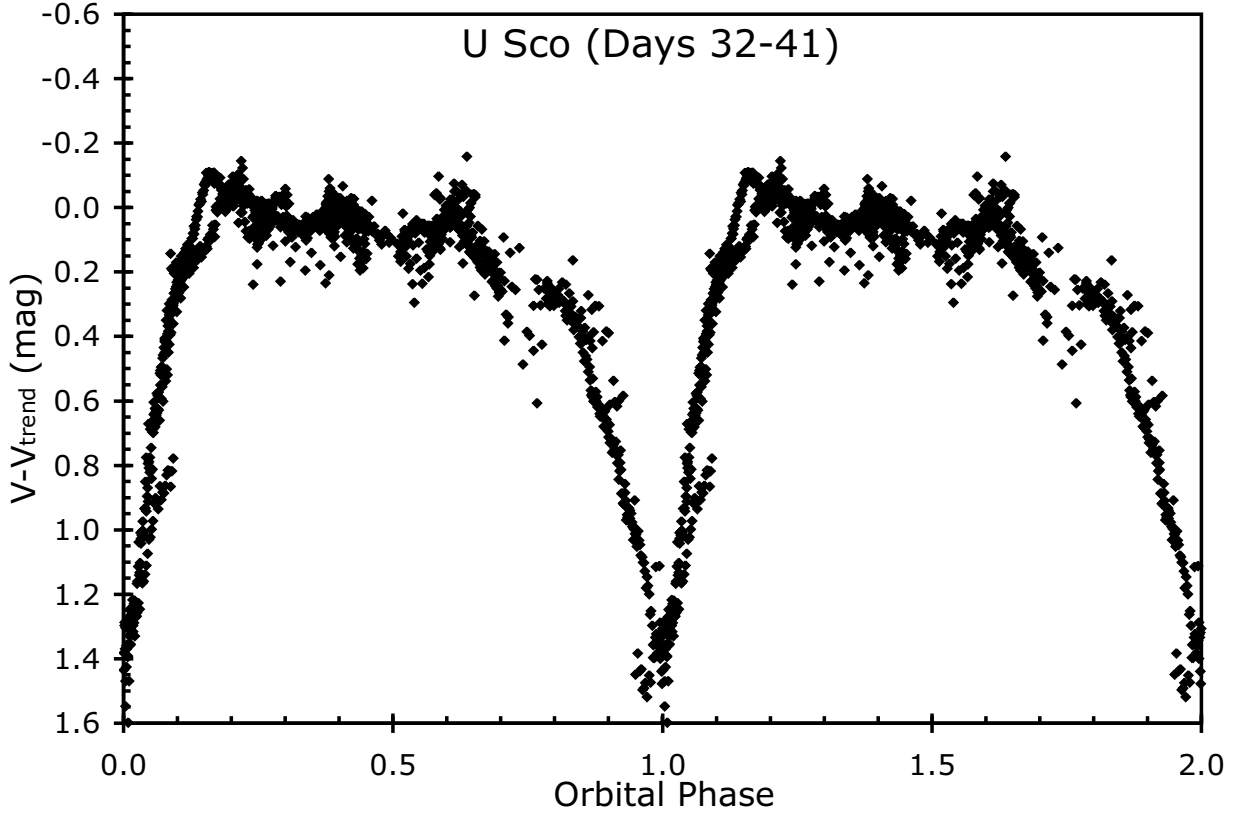}
\caption{
Days 32-41.  The eclipses get very deep, with the 1.4 mag amplitude implying that the companion star (2.66 AU in radius) is covering up 75\% of the system's light.  Eclipse mapping shows that the central light source is consistent with a centrally-bright disk with radius around 2.2 R$_{\odot}$, which is similar to the quiescent state.  However, some of the egresses are a bit wider than in quiescence, indicating some residual material outside the stabilizing disk.  The system has suddenly stopped showing the secondary eclipse, despite its prominence from days 26-32.  The light curve shows two asymmetries, a steady fading by a quarter of a magnitude from phase 0.25 to 0.75, and a slow ingress relative to the egress.  Both asymmetries can be explained by material in the accretion stream and near the usual hot spot position providing occultation of the inner light source as well as a bright inner edge best visible just after eclipse.}
\end{figure}

\clearpage
\begin{figure}
\epsscale{1.0}
\plotone{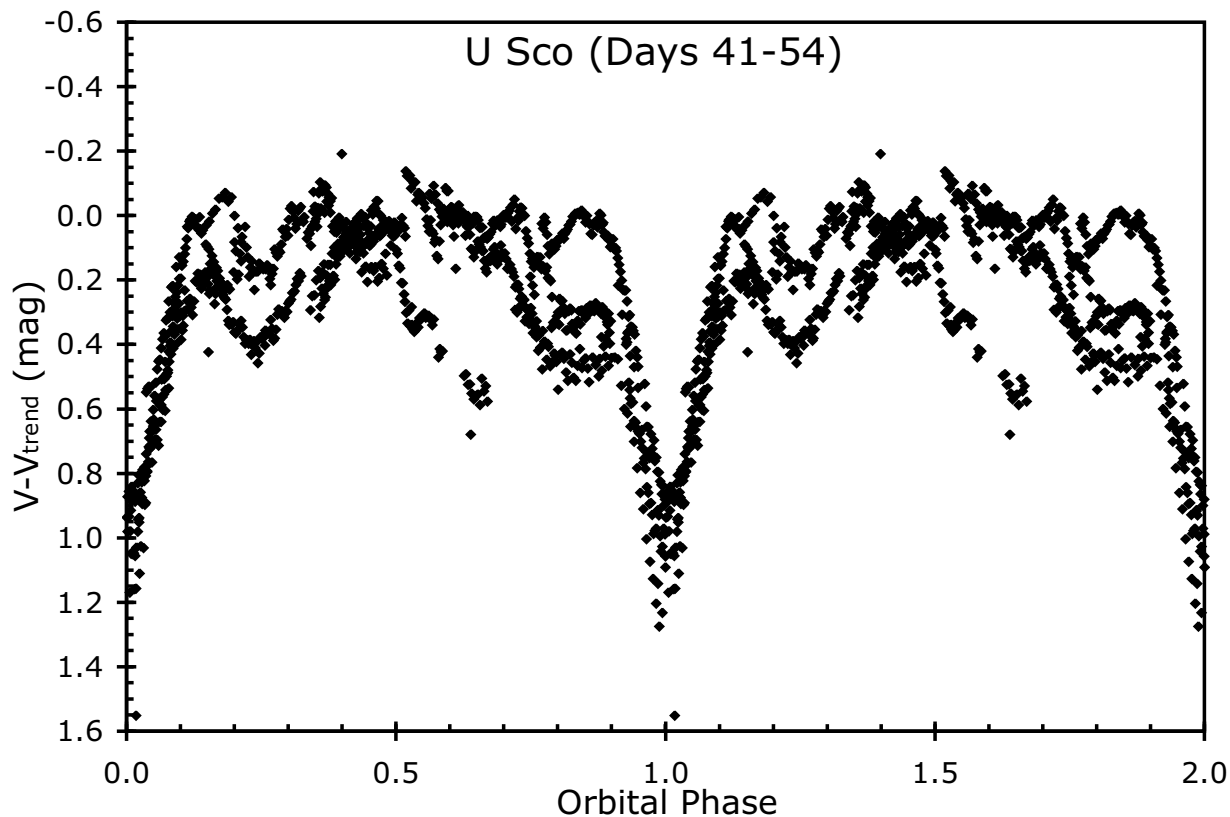}
\caption{
Days 41-54.  The stunning change is that the out-of-eclipse intervals show deep dips that vary greatly from orbit to orbit.  These dips get as deep as 0.6 mag with typical durations of 0.2 in phase.  (A non-phased light curve of the dips can be seen in Figure 14.)  This phenomenon has no precedent in novae at any time, and here we propose that these dips are analogous to the dips in low mass X-ray binaries.  The secondary eclipse and the light curve asymmetries have disappeared, although this could well be confused by the dips.  The primary eclipses have a depth of 1.0 mag, while the duration has shortened greatly to 0.20 in phase.  In contrast to the previous days, the ingress is substantially faster than the egress.}
\end{figure}

\clearpage
\begin{figure}
\epsscale{1.0}
\plotone{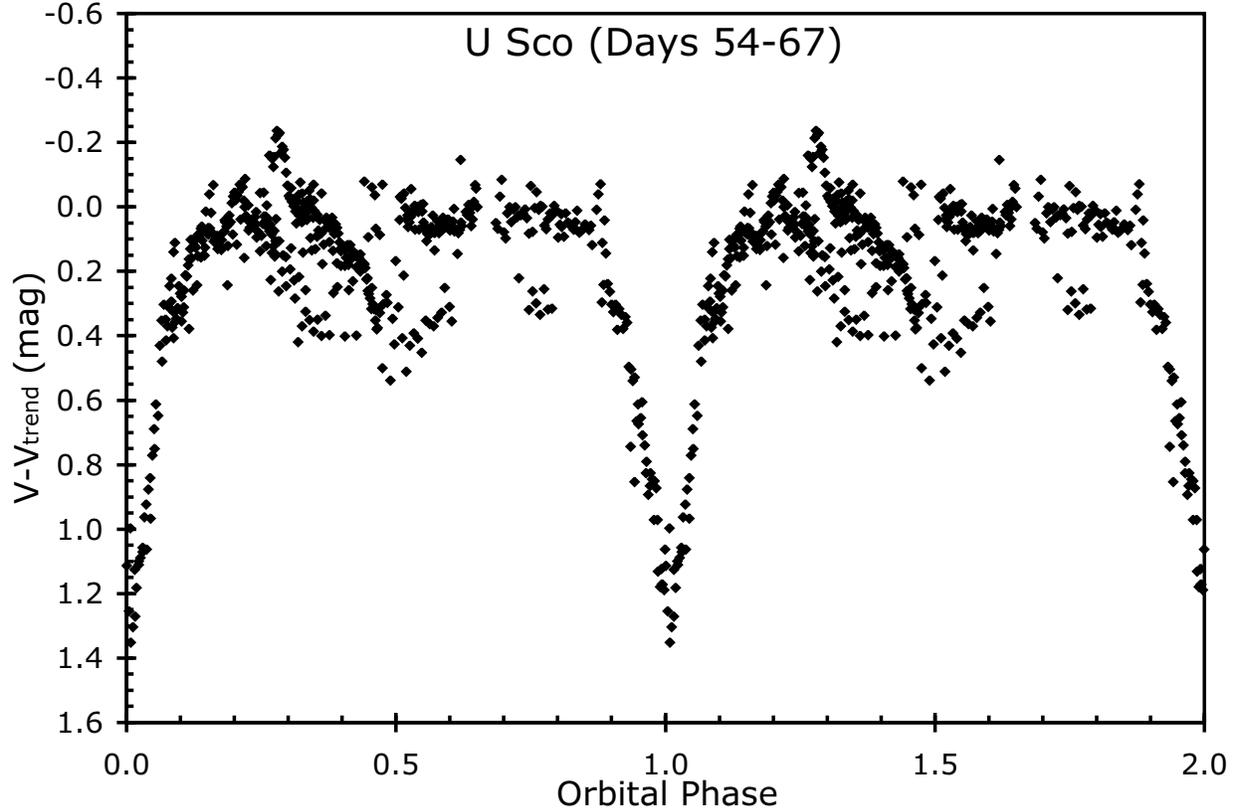}
\caption{
Days 54-67.  This light curve has similar properties to the one during the interval from days 41-54, but we have kept them separate in two figures so that some of the runs from individual nights can be distinguished.  The eclipse looks slightly deeper (1.1 mag), somewhat longer in duration (0.27 in phase), and nearly symmetric in shape.  The key feature of this light curve is the presence of deep and broad dips that continue to occur, apparently randomly in phase.}
\end{figure}

\clearpage
\begin{figure}
\epsscale{1.0}
\plotone{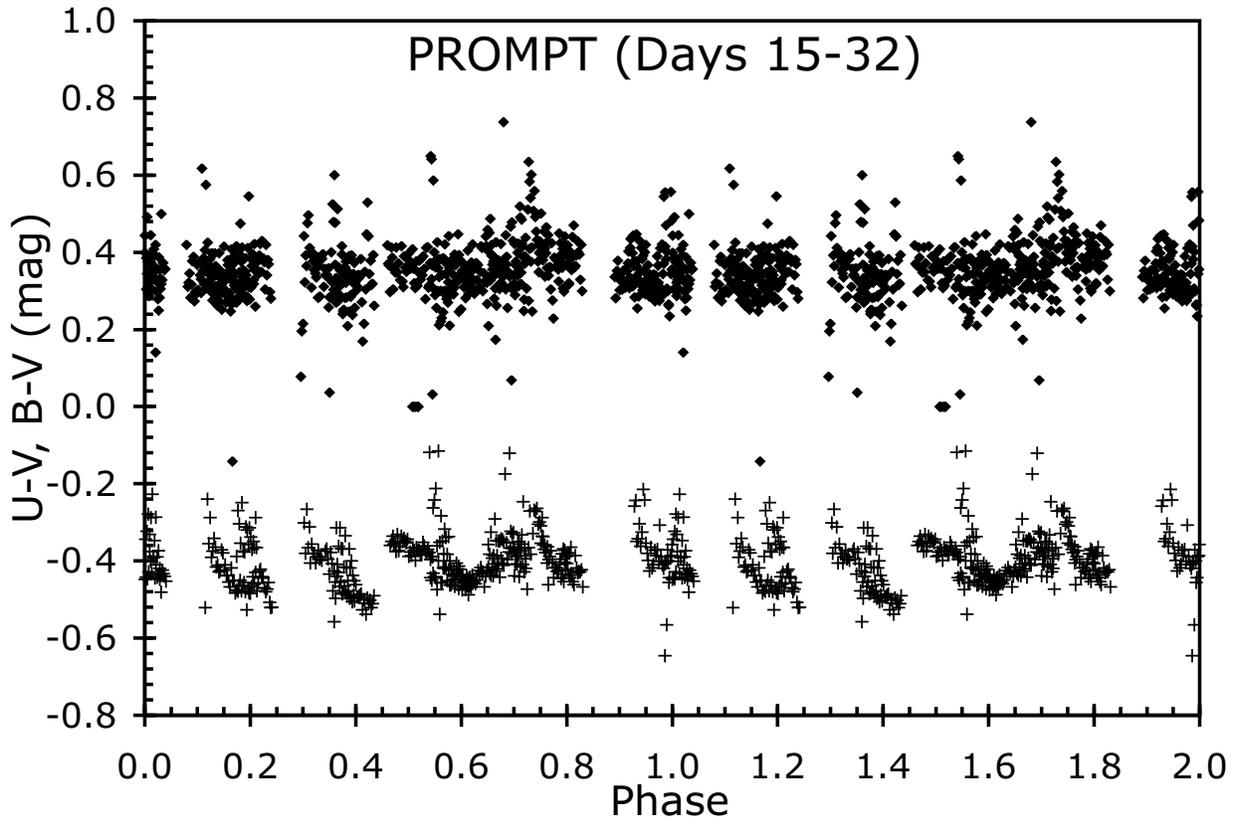}
\caption{
U-V and B-V color curves.  This shows the PROMPT data for days 15-32 with the first plateau.  The U-V values are shown as crosses, while the B-V are shown as filled diamonds.  The system brightness (see Figures 4-6) varies greatly due to primary eclipses, secondary eclipses, and out of eclipse asymmetries.  Yet through all this, the U-V and B-V colors remain constant.}
\end{figure}

\clearpage
\begin{figure}
\epsscale{1.0}
\plotone{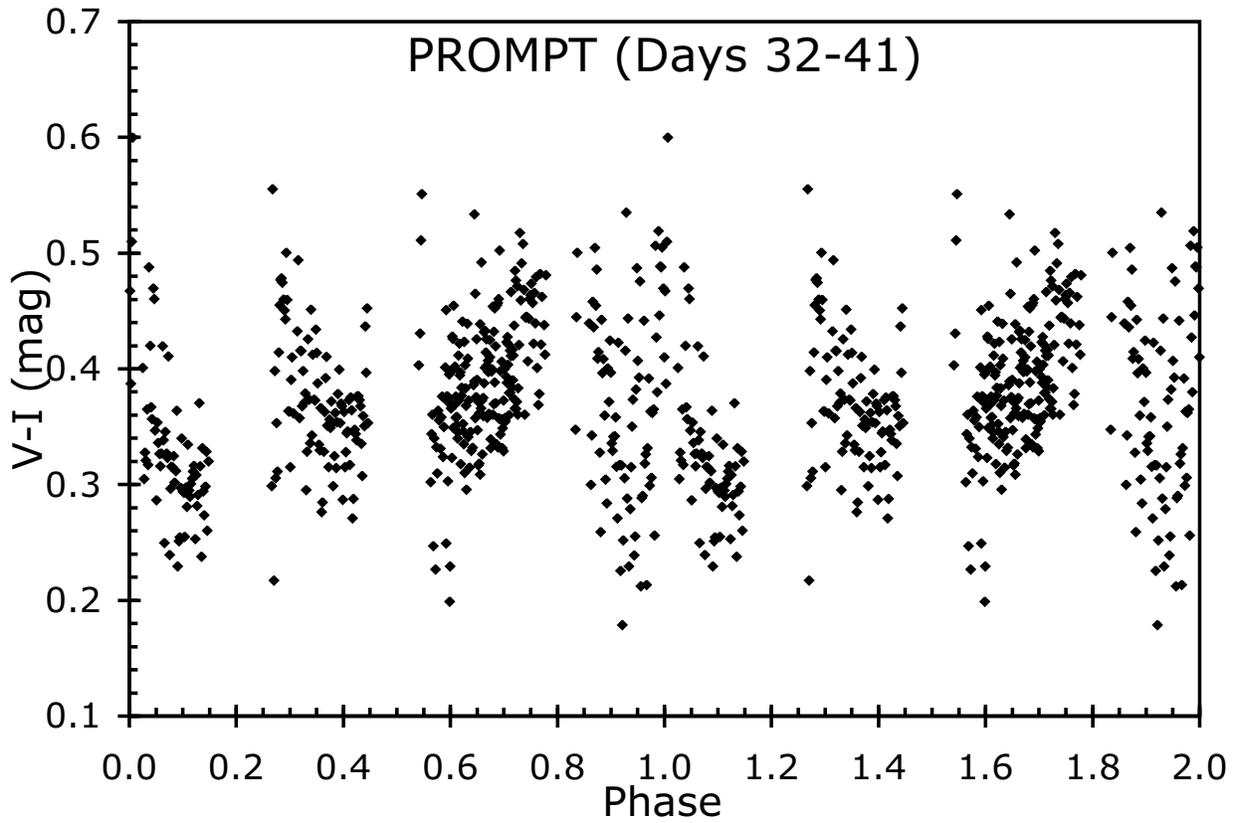}
\caption{
V-I color curve.  For days 32-41, the brightness is fast falling from the first plateau and shows deep eclipses (see Figure 7).  Through all this, the V-I color remains constant.  We interpret this as evidence that most of the light in the system comes from Thompson scattering from a single central source.}
\end{figure}

\clearpage
\begin{figure}
\epsscale{1.0}
\plotone{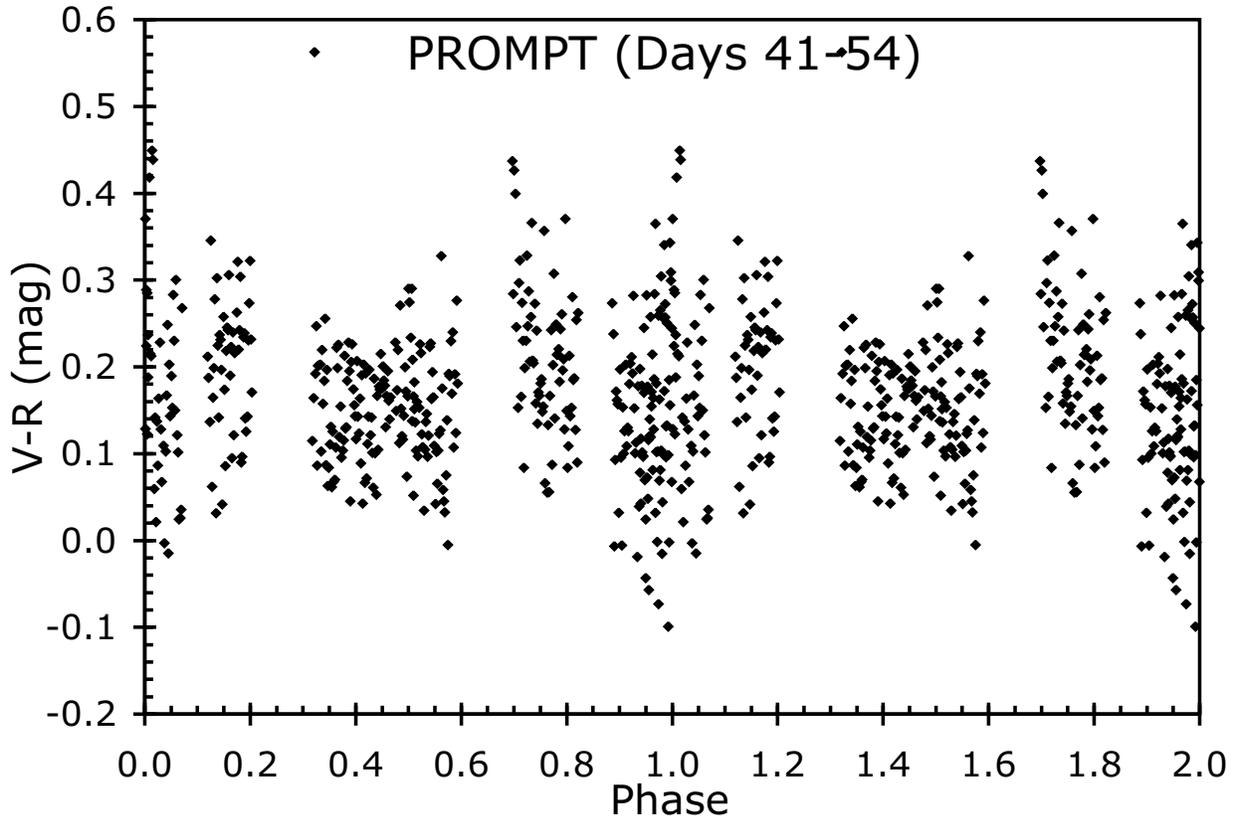}
\caption{
V-R color curve.  Again, we see that the color remains constant through eclipses (compare with Figure 8).  It also remains constant through the dips.  The observed V-R colors have a RMS scatter of 0.09 mag, which is moderately close to the median one-sigma error bars of 0.05 mag (with 18\% of the points having error bars larger than 0.10 mag), so most of the observed scatter is likely just from the normal measurement errors.  This constancy of color is true throughout the entire eruption, for all the colors.}
\end{figure}

\clearpage
\begin{figure}
\epsscale{1.0}
\plotone{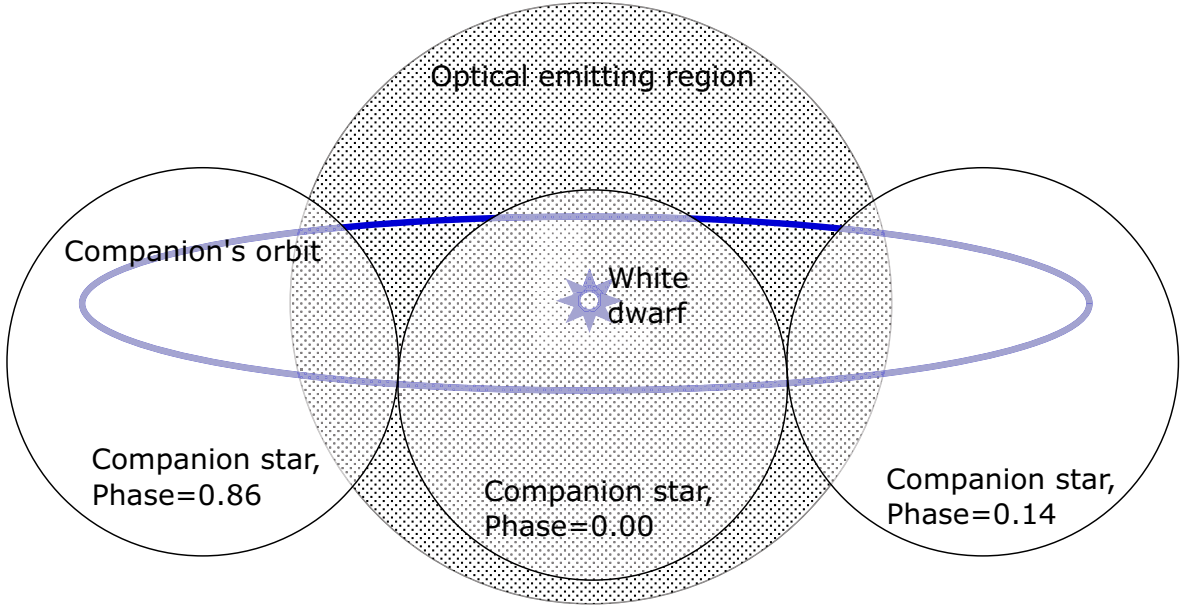}
\caption{
Eclipse mapping.  U Sco is viewed with its orbital plane near edge-on ($\approx80\degr$ inclination), with the separation between the star centers equal to 6.87 R$_{\odot}$ and the radius of the secondary equal to 2.66 R$_{\odot}$ (Hachisu et al. 2000a; 2000b).  The simplest model for eclipse mapping has the companion star occulting a uniform circular source with a sharp edge, and this model provides a good explanation for the eclipse light curve for days 15-26.  The figure illustrates to scale the companion star (the translucent circles) at three different orbital phases, the white dwarf (the central small star icon), the orbit of the companion (the large ellipse), and the optical emitting region (the large shaded circle).  Early in the eruption, the effective radius of the optical source is 4.1 R$_{\odot}$, with this decreasing somewhat until day 26.  This spherical source is the photosphere and wind produced by the supersoft X-ray source and cuts off just inside the orbit of the secondary star.  For days 26-41, the simplest model (and all variants involving radial symmetry for the optical source) cannot account for the wide eclipses.  Indeed, for these days, nearly all of the optical flux is eclipsed with the companion star at just three phases (0.86, 0.00, and 0.14) for which there is no overlap.  With this, around the end of the plateau on its decline, nearly all of the optical light must be coming from near the orbital plane.  Indeed, with eclipse mapping for a disk model, the templates for days 26-32 and 32-41 are shown to arise from a rim-brightened disk with radius 3.4 R$_{\odot}$.}
\end{figure}

\clearpage
\begin{figure}
\epsscale{1.0}
\plotone{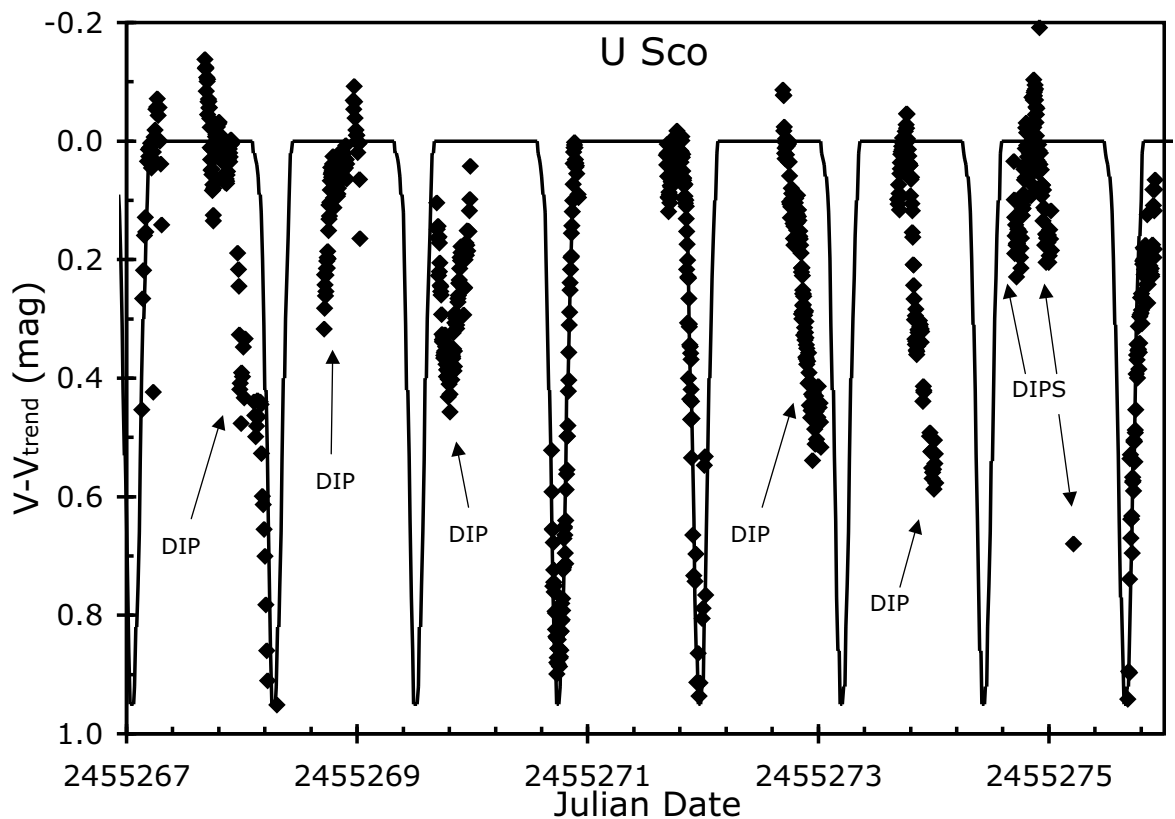}
\caption{
Optical dips.  From days 41-61, the detrended light curve shows deep dips at apparently random phases, with these certainly not being associated with the regular primary eclipses caused by the secondary star.  This plot shows the light curve template (from Table 6) plus all our observed magnitudes from JD 2455267.0 to 2455276.0 (roughly days 43-52).  The optical light source is fairly small and centered on the white dwarf (as demonstrated by the depth and timing of the primary eclipses), so the dips can only be eclipses caused by occulters spread around the white dwarf.  U Sco has an inclination of roughly 80-84$\degr$, so our line of sight to the bright central source is just skimming over the top of the disk, such that a small increase in the height of the disk rim will dim the entire system for a small range of phases.  Raised rims are expected during the re-establishment of the disk, as the accretion stream moves around.  This optical dipping is unique among novae, although low mass X-ray binary systems with neutron stars are often seen to have X-ray dips.}
\end{figure}

\clearpage
\begin{figure}
\epsscale{1.0}
\plotone{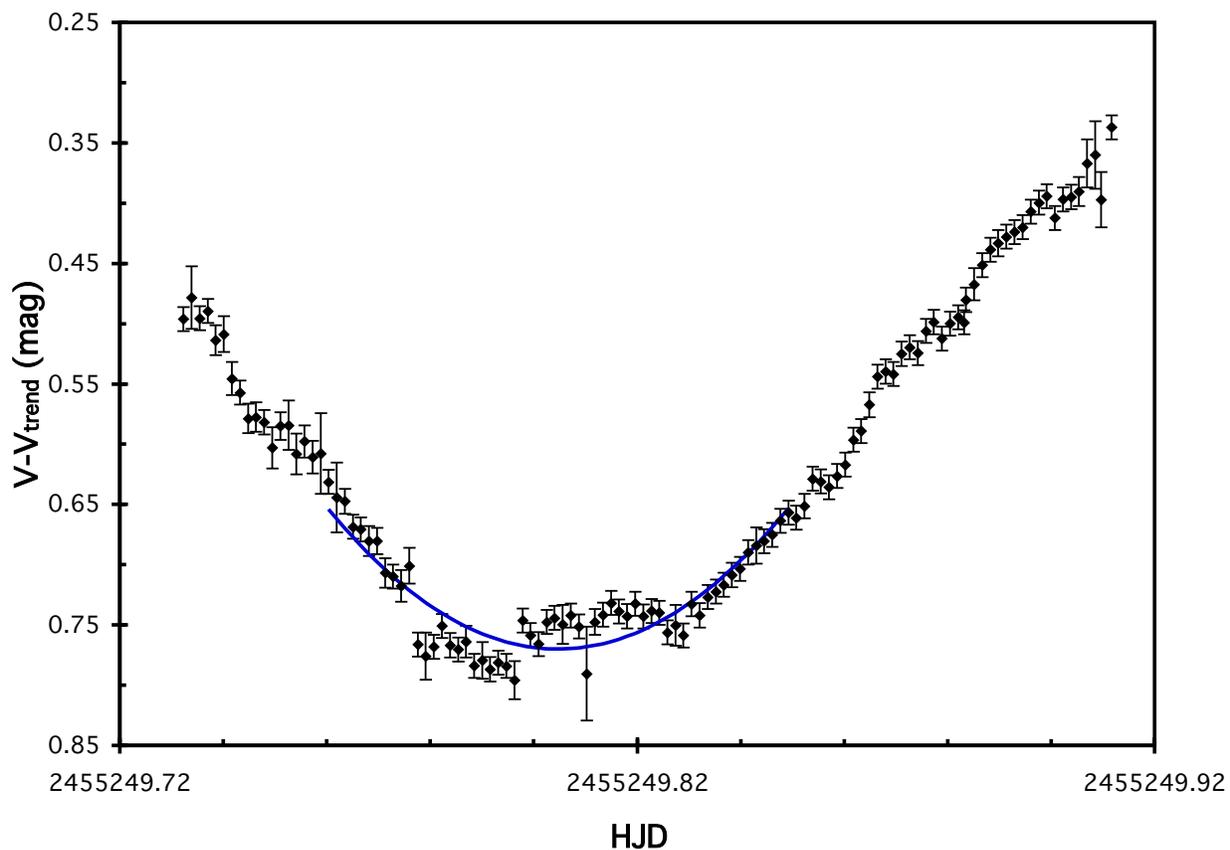}
\caption{
Eclipse on day 25.  This light curve comes entirely from a single run by one observer (Oksanen).  We see an eclipse that is certainly not `V-shaped', but rather looks fairly flat across the bottom.  The duration of the apparent totality is 1.23 hours.  The best fit parabola is displayed over the fit range, with this giving a minimum time of HJD=2455249.8047$\pm$0.0008 and a minimum of $0.770\pm0.002$ mag.  The scatter of the individual magnitudes around the best fit parabola (or any other appropriate curve) is larger than the error bars, likely as a result of the variability and non-uniformity of the light source being eclipsed, with the result that the time of minimum has a systematic error larger than the uncertainty deriving from the measurement errors alone.}
\end{figure}

\clearpage
\begin{figure}
\epsscale{1.0}
\plotone{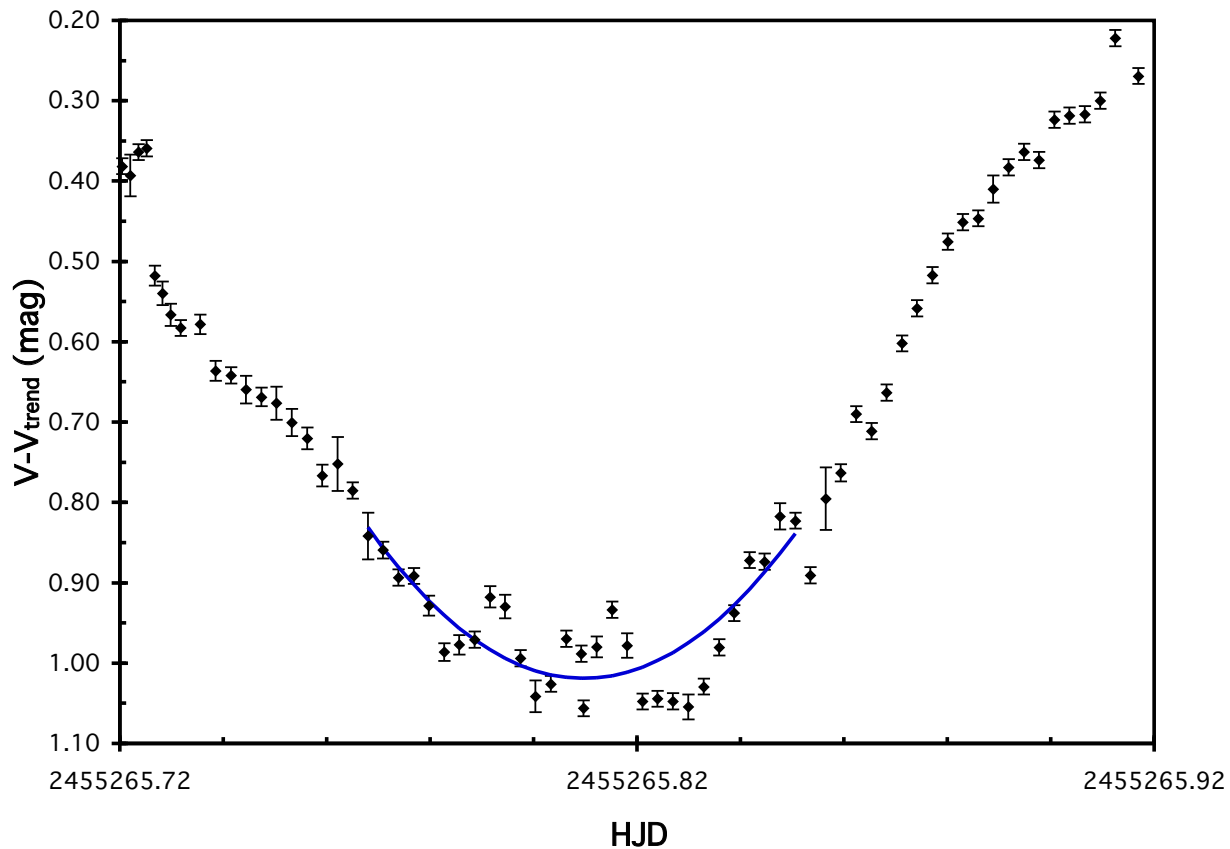}
\caption{
Eclipse on day 41.  Again, this light curve (entirely from Oksanen) looks effectively flat across the bottom rather than `V-shaped'.  The duration of the apparent totality is 1.27 hours.  The photometric scatter during totality is substantially larger than our photometric errors, and points to a systematic variation intrinsic to U Sco.  The best fit parabola is calculated by a chi-square minimization where the total uncertainty for each point is the addition in quadrature of the measurement error and some systematic uncertainty selected so that the reduced chi-square is unity.  With this procedure, the size of the one-sigma error bars for the minimum time (for the period range over which the chi-square is within 1.00 of the minimum chi-square) will account for the intrinsic variations.  The parabola is displayed over the time range for the fit, with best fit parameters of HJD=2455265.8097$\pm$0.0015 and minimum $V-V_{trend}=1.019\pm0.007$ mag.}
\end{figure}

\clearpage
\begin{figure}
\epsscale{1.0}
\plotone{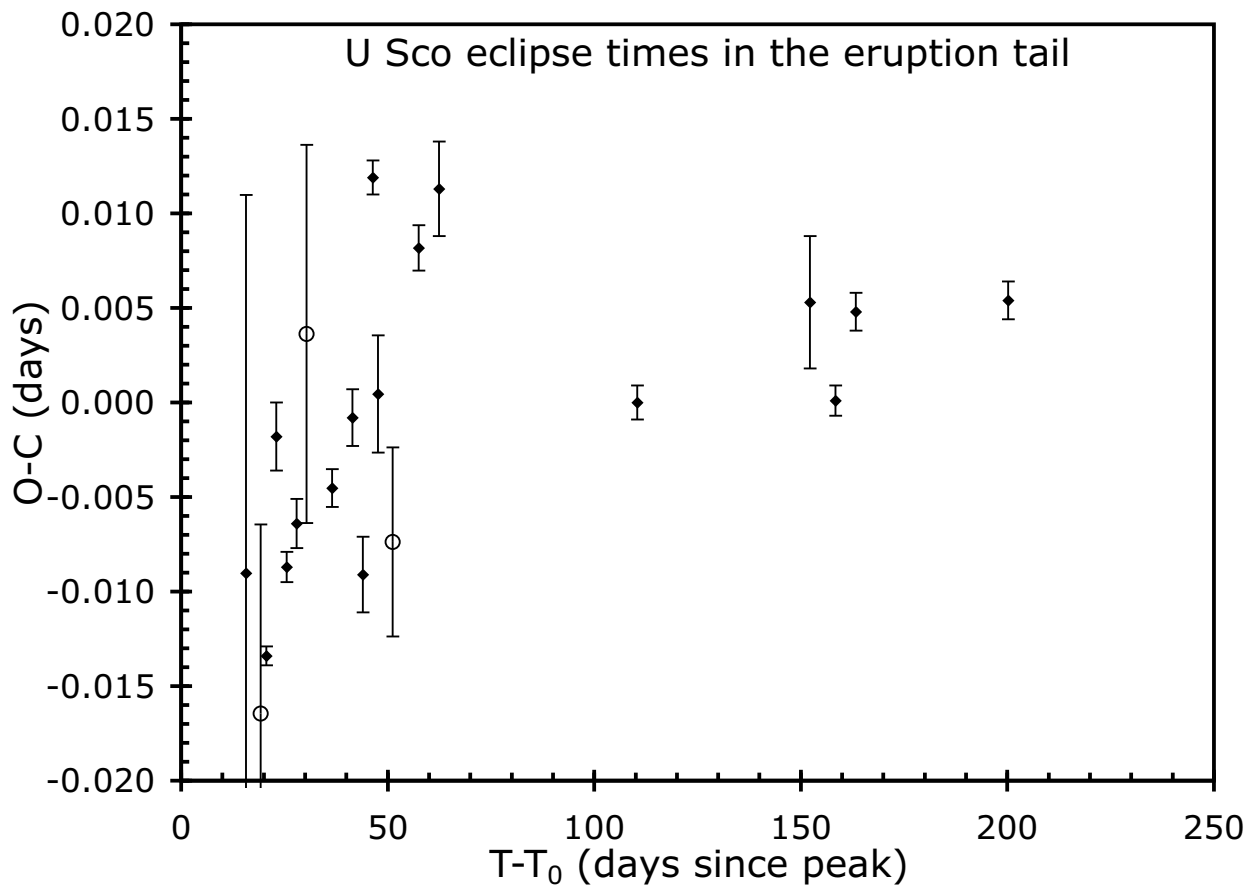}
\caption{
Eclipse times.  The vertical axis shows the O-C values for the observed eclipse times relative to the linear ephemeris in Equation 1.  The filled diamonds are our eclipse times for the 2010 eruption, while the empty circles are for times during the 1999 eruption.  This figure shows that the eclipse times have a systematic and significant offset from zero, and this offset changes roughly linearly time time until the end of the eruption.  The eclipse times throughout the eruption vary about any smooth curve (linear or otherwise) with a scatter larger than the quoted uncertainties.  This orbit-to-orbit variation is intrinsic to U Sco, and is another manifestation of the variability and fine structure of the eclipsed region.}
\end{figure}

\end{document}